\documentclass[twocolumn,showpacs,amsfonts,aps,prc,floatfix,
]{revtex4} 

\usepackage{amsmath}
\usepackage{bm}
\usepackage{graphicx}

\newcommand{\be}[1]{\begin{equation}\label{#1}}
\newcommand{\ee}{\end{equation}}

\usepackage{color}

\begin{document}


\title{Relativistic Dissipative Hydrodynamic Equations
at the Second Order
for Multi-Component Systems with Multiple Conserved Currents} 
\date{\today}

\author{Akihiko Monnai}
\email{monnai@nt.phys.s.u-tokyo.ac.jp}
\affiliation{Department of Physics, The University of Tokyo,
Tokyo 113-0033, Japan}

\author{Tetsufumi Hirano}
\email{hirano@phys.s.u-tokyo.ac.jp}
\affiliation{Department of Physics, The University of Tokyo,
Tokyo 113-0033, Japan}

\begin{abstract}
We derive the second order hydrodynamic equations 
for the relativistic  system of multi-components with multiple conserved
currents by generalizing the Israel-Stewart theory and Grad's moment method.
We find that, in addition to the conventional
moment equations,
extra moment equations associated with conserved currents 
should be introduced 
to consistently match the number of equations
with that of unknowns and to satisfy the Onsager reciprocal relations.
Consistent expansion of the entropy current
leads to constitutive equations which 
involve the terms not appearing in the original Israel-Stewart theory
even in the single component limit.
We also find several terms which exhibit thermal diffusion
such as Soret and Dufour effects.
We finally compare our results with those of other existing formalisms.
\end{abstract}

\maketitle

\section{Introduction}
\label{sec1}
\vspace*{-2mm}

Hydrodynamics, which is grounded on
conservation laws under local equilibrium conditions, 
is widely used
in general physics.
Its relativistic version 
taking account of irreversible
processes
was initiated by Eckart \cite{Eckart:1940te} many years ago.
Later, Landau \cite{Landau} applied relativistic hydrodynamics
to multi-particle production in hadron-hadron collisions
in cosmic ray events.
Just after the historical work,
the applicability of 
relativistic hydrodynamics by Landau
was examined in terms of
quantum field theories \cite{namikiiso}.
However, the theories by Eckart \cite{Eckart:1940te} and by Landau
\cite{LandauLifshitz}
share a common problem
that dissipative perturbation
propagates at infinite
speed \cite{HiscockLindblom}, which is
obviously incompatible with
the concept of causality in 
the relativistic theory.
The problem is originated from
rather \textit{phenomenological} constitutive equations
for dissipative currents:
Instantaneous responses
to thermodynamic forces,
which are usually assumed
in non-relativistic irreversible
processes such as Fourier's law and Newton's law,
lack the relaxation of the dissipative currents.
The theory is called first order theory
when the dissipative currents are proportional to
thermodynamic forces since
the entropy current in ideal hydrodynamics
is corrected by
the linear terms of dissipative currents.
On the other hand,
second order theory, in which
the entropy current
has quadratic terms of dissipative currents,
leads to the relaxation terms
for dissipative currents and therefore
can satisfy the causality.
So far, a wide variety of second order theories
\cite{Israel:1976tn,Israel:1979wp,
Muller,Geroch:1990bw,Carter:1991,
Prakash:1993bt,Ottinger,
Muronga,Koide:2006ef,Tsumura:2007ji,Baier:2007ix,Bhattacharyya:2008jc,Natsuume:2007ty,Lublinsky:2009kv,PeraltaRamos:2009kg, Betz:2009zz}
have been proposed
as relativistic theories
of irreversible processes.
The expression of
constitutive equations, however,
varies among theories.

In this paper,
we derive relativistic dissipative hydrodynamic
equations
at the second order with
multi-components as well as multiple conserved currents.
One finds several approaches
to obtain constitutive equations
for relativistic systems in the literature.
First of all, to clarify the aim of this paper,
we categorize the systems into four classes
according to the number of components and their
interaction:
(a) single component with binary collisions,
(b) single component with elastic and inelastic collisions,
(c) multi-components with binary collisions and
(d) multi-components with elastic and inelastic collisions.
In all the classes above,
one of the macroscopic equations to be solved
is the energy-momentum conservation.
The typical situation, which can be found in many textbooks
of kinetic theory or non-equilibrium statistical physics,
belongs to
the first class (a), in which the number of particles
is conserved during evolution.
In addition to the energy-momentum conservation,
one needs to solve the continuity equation of the
number of particles.
In the second class (b), the number
of particles is determined locally by temperature
and is not necessary conserved due to inelastic processes
during evolution
under local thermal and chemical equilibrium.
Then, one solves the energy-momentum conservation only
in this case.
In the third class (c),
the number of each component is conserved
due to binary collisions. Thus,
the number conservation for each component
as well as the energy-momentum conservation
are solved simultaneously.
In the fourth class (d),
which we will discuss in this paper,
the number of each component may not be conserved
due to inelastic collisions or chemical reactions.
However, there can exist several
conserved numbers due to symmetry of Lagrangian
under some continuous transformation.
Instead of the number conservation for each component,
one needs to solve the continuity 
equations of \textit{conserved charges}
together with energy-momentum conservation.
Note that the number of components does not need to coincide
with the number of conserved currents.
To our best knowledge, no systematic investigation
is available for the class (d) above
even though it is the most
important situation in ultra-relativistic systems
at very high temperature
in which particle creation and annihilation
take place frequently.

The importance of relativistic hydrodynamics \cite{reviews1}
has been increasing after the discovery of
the ``perfect fluid" quark-gluon plasma (QGP)
in Au+Au collisions at $\sqrt{s_{NN}} = 200$ GeV
at Relativistic Heavy Ion Collider (RHIC)
in Brookhaven National Laboratory (BNL) \cite{experiments}.
This fact has been quantified by reproducing,
within ideal hydrodynamic models \cite{HKH, Teaney:2000cw, Hiranov2eta},
particle spectra
as well as elliptic flow coefficients $v_2$
(ellipticity of radial flow in momentum space \cite{Ollitrault})
as functions of centrality, transverse momentum
and pseudorapidity from
experimental data \cite{STARv2, PHENIXv2, PHOBOSv2}.
Since ideal hydrodynamic models are approximation
in the sense that all non-equilibrium processes are omitted,
our next step should be to include small viscosity
to capture the correct physics.
The reasonable agreement of ideal hydrodynamic results
with experimental data suggests that the system is not
so far from equilibrium, \textit{i.e.},
viscous hydrodynamic models can be justified for
the QGP at RHIC energies.
In Large Hadron Collider (LHC) experiments
which have just begun \cite{Collaboration:2009dt}
and are planned to eventually reach $\sqrt{s_{NN}} = 5.5$ TeV
in the heavy ion program,
viscous hydrodynamic models will become even more important
in quantifying the properties of the hot QCD matter
and examining the applicability of hydrodynamic models.
It should be noted here that hydrodynamics
and a hydrodynamic model are different concepts
and thus are to be distinguished;
the former is a general macroscopic theory that
describes strongly-coupled relativistic systems,
while the latter is a specific model based on
hydrodynamics that describes the phenomena of interest,
namely relativistic heavy ion collisions.
It is essential that viscous hydrodynamics be
established before constructing any realistic models
for heavy ion collisions.

We aim to develop the formalisms
of relativistic dissipative hydrodynamics
for multi-component systems with multi-conserved currents by
determining the distortion of distribution functions and then 
constraining the constitutive equations for the dissipative currents.
The discussion for multi-component systems was not recognized well
in the context of highly relativistic system where
particle creation and annihilation take place,
but it turned out
to be far from trivial \cite{Monnai:2009ad}.
Multi-component hydrodynamics
is important in developing dissipative hydrodynamic models
for the hot QCD matter, 
which also is a multi-component system.
We also consider systems with multiple conserved currents
because one should be able to introduce
more than one conserved charge
such as baryon number and strangeness to the system.

This paper is organized as follows.
In Sec.~\ref{sec2}, we show how to consistently formulate
the second order constitutive equations for multi-component systems.
The systems with multiple conserved currents
are considered.
In Sec.~\ref{sec3}, we discuss the correspondences
between our results and several different existing
second order equations.
The conclusion will be given in Sec.~\ref{sec4}.
The Minkowski metric $g^{\mu \nu} = \mathrm{diag}(+,-,-,-)$
and the natural unit $c = \hbar = k_B = 1$ are used throughout this paper.

\section{Derivation of Second-Order Viscous Hydrodynamic Equations}
\label{sec2}

We derive macroscopic dissipative hydrodynamic equations
in multi-component systems with multiple conserved currents
by extending the Israel-Stewart second order theory \cite{Israel:1979wp}.
We consider the third moment of the distribution function $f^i$
and constrain its derivative from the second law of thermodynamics.
We also discuss additional moment equations which do not appear
in the conventional Israel-Stewart formalism to consistently
describe multi-component systems.

We find several non-trivialities of multi-component systems
in the course of the formulation.
Firstly, the thermodynamic stability conditions,
which ensure that the system is in maximum entropy state
in terms of dissipative currents,
have to be employed after the constitutive equations
for dissipative currents are obtained,
because the number of the constitutive equations
and that of the dissipative currents would not match
if the conditions were considered beforehand.
Secondly, the second law of thermodynamics requires
a specific tensor structure for moment expansion
of the distortion of the distribution $\delta f^i$
in a multi-component system,
which justifies the result of Ref.~\cite{Monnai:2009ad}.
Thirdly, if the system has conserved currents
we need to consider new moment equations to
consistently formulate multi-component relativistic dissipative hydrodynamics.
These equations also allow us to determine all the dissipative
currents in an arbitrary frame, which the conventional
Israel-Stewart theory would not do. 

The existence of multiple conserved currents also brings uncertainties 
to the conventional Grad's 14-moment method
since it is no longer applicable when more than 
14 dissipative currents are present.
In this paper we consider systems with conservations
based on quantum numbers, such as baryon number,
strangeness and isospin.
In other words, inelastic scattering
and chemical interactions are present.
It should be noted again that the number of conserved currents
and that of particle species
in the system are generally different
when inelastic processes are present.
We propose a generalized moment
method based on Onsager reciprocal relations \cite{Onsager}
to describe such systems without ambiguity.

\subsection{Extended Second Order Israel-Stewart Theory for Multi-Component Systems with Multiple Conserved Currents}
\label{subsec:eis}

We would like to introduce thermodynamic quantities in the tensor decompositions of the energy-momentum tensor and the conserved currents.
In the ideal relativistic hydrodynamics
the energy-momentum tensor and the conserved currents are
expressed as
\begin{eqnarray}
\label{eq:em_eq}
T_0 ^{\mu \nu} &=& e_0 u^\mu u^\nu - P_0 \Delta ^{\mu \nu} , \\
\label{eq:n_eq}
N_{J0} ^{\mu} &=& n_{J0} u^{\mu},
\end{eqnarray}
where the index $J \ (J=1, \cdots,N)$ denotes different types
of conserved currents.
Here $N$ is the number of conserved currents.
$u^\mu$ is the four velocity normalized as $u^\mu u_\mu = 1$.
$\Delta^{\mu \nu} = g^{\mu \nu} - u^\mu u^\nu$
is the projection operator.
$e_0 = u_\mu T_0^{\mu \nu} u_\nu$,
$P_0 = -\frac{1}{3} \Delta_{\mu \nu} T_0^{\mu \nu}$
and $n_{J0} = u_\mu N_{J0}^\mu$
denote the energy density, the hydrostatic pressure and
the charge number density of the $J$-th conserved current, respectively.
The number of unknowns here is $5+N$
because one has $e_0$(1 unknown), $P_0$(1 unknown), $n_{J0}$ ($N$ unknowns)
and $u^\mu$(3 unknowns).
On the other hand there are only $4+N$ equations
from the energy-momentum conservation and
the charge number conservations to describe the system:
\begin{eqnarray}
\label{eq:conservation}
\partial _\nu T^{\mu \nu} &=& \partial _\nu T_{0}^{\mu \nu} \ = \ 0 , \\
\label{eq:conservation2}
\partial _\mu N_J^{\mu} &=& \partial _\mu N_{J0}^{\mu} \ = \ 0 .
\end{eqnarray}
Therefore, one needs to introduce the equation of state $P_0 = P_0(e_0, \{n_{J0}\})$ from microscopic physics to completely determine the space-time evolution of the system.

On the other hand,
the energy-momentum tensor and the conserved currents
in relativistic dissipative hydrodynamics are tensor-decomposed into
\begin{eqnarray}
\label{eq:em_eqnoneq}
T ^{\mu \nu} &=& (e_0+\delta e) u^\mu u^\nu - (P_0+\Pi ) \Delta ^{\mu \nu} \nonumber \\
&+& 2W^{( \mu} u^{\nu )} + \pi^{\mu \nu} , \\
\label{eq:n_eqnoneq}
N_J ^{\mu} &=& (n_{J0}+ \delta n_J) u^{\mu} + V_J^{\mu}.
\end{eqnarray}
We denote the dissipative parts of the above equations
as $\delta T^{\mu \nu} = T^{\mu \nu} - T^{\mu \nu}_0$
and $\delta N_J^\mu = N_J^\mu - N_{J0}^\mu$.
Then $\Pi = -\frac{1}{3} \Delta_{\mu \nu} \delta T^{\mu \nu}$
is the bulk pressure,
$W^\mu = \Delta ^\mu _{\ \alpha} \delta T^{\alpha \beta} u_\beta $
the energy current,
$\pi ^{\mu \nu} = \delta T ^{\langle \mu \nu \rangle}
= [ \frac{1}{2} (\Delta ^\mu _{\ \alpha} \Delta ^\nu _{\ \beta}
+ \Delta ^\mu _{\ \beta} \Delta ^\nu _{\ \alpha})
- \frac{1}{3}\Delta^{\mu \nu} \Delta_{\alpha \beta}]
\delta T^{\alpha \beta} $
the shear stress tensor
and $V_J^\mu = \Delta ^\mu _{\ \nu} \delta N^\nu_J$
the charge current of the $J$-th conserved current.
Round bracket for Lorentz indices denotes
symmetrization as
$A^{( \mu} B^{\nu )} = \frac{1}{2}(A^\mu B^{\nu}+B^\mu A^{\nu})$. 
$\delta e = u_\mu \delta T^{\mu \nu} u_\nu$
and $\delta n_{J} = u_\mu \delta N_J^\mu$
are the distortion of the energy density
and of the $J$-th charge density, respectively.
They actually vanish because the stability conditions
need to be employed.
The details can be found in 
Appendix of Ref.~\cite{Monnai:2009ad}.
However, we have to keep these quantities
for the moment to correctly count the number of unknowns.

The energy-momentum conservation
and the charge number conservations 
in non-equilibrium systems are tensor-decomposed into
\begin{eqnarray}
\label{eq:em_par}
D (e_0 + \delta e) &=& 
- (e_0 + \delta e + P_0 + \Pi ) \nabla _\mu u^\mu \nonumber \\
&+& 2 W^\mu D u_\mu - \nabla _ \mu W^\mu \nonumber \\
&+& \pi^{\mu \nu} \nabla _{\langle \mu} u_{\nu \rangle} , \\
\label{eq:em_ort}
(e_0 + \delta e + P_0 + \Pi ) D u^\mu &=& \nabla ^\mu (P_0 + \Pi ) 
- W^\mu \nabla _\nu u^\nu \nonumber \\
&-& \Delta^{\mu \nu} D W_\nu - W^\nu \nabla _\nu u^\mu \nonumber \\
&+& \pi^{\mu \nu} D u_\nu - \Delta ^{\mu \nu} \nabla ^\rho \pi_{\nu \rho},\\
\label{eq:n_par}
D (n_{J0} + \delta n_J) &=& - (n_{J0} + \delta n_J) \nabla _\mu u^\mu \nonumber \\
&-& \nabla _\mu V_J^\mu + V_J^\mu D u_\mu ,
\end{eqnarray}
where the time-like and the space-like derivatives
are defined as $D = u^\mu \partial _\mu$
and $\nabla _\mu = \Delta _{\mu \nu} \partial ^\nu$, respectively.

There are $10+4N$ additional macroscopic variables $\Pi$,
$\delta e$, $W^\mu$,  $\pi^{\mu \nu}$, $\delta n_J$
and $V_J^\mu$, which originate from $10$ independent components of
$\delta T^{\mu \nu}$ and $4N$ independent components
of $\delta N_J^{\mu}$.
Therefore one needs to introduce \textit{constitutive equations}
for these dissipative currents
to completely describe the system.
The relativistic Navier-Stokes equations of 
the dissipative currents can be obtained 
by considering the law of increasing entropy.
The energy momentum tensor $T^{\mu \nu}$,
the conserved currents $N_J^\mu$
and the entropy current $s^\mu$
are expressed in kinetic theory
with the microscopic phase-space distribution $f^i$ as
\begin{eqnarray}
\label{eq:second}
T^{\mu \nu} &=& \sum _i \int \frac{g_i d^3 p}{(2\pi )^3 E_i}
p_i^\mu p_i^\nu f^i , \\
\label{eq:first}
N_J^{\mu} &=& \sum _i \int \frac{q_i^J g_i d^3 p}{(2\pi )^3 E_i}
p_i^\mu f^i , \\
\label{eq:entropycurrent}
s^\mu &=& - \sum_i \int \frac{g_i d^3 p}{(2\pi )^3 E_i}
p_i ^\mu \phi (f^i) ,
\end{eqnarray}
where $g_i$ is the degeneracy and $q_i^J$
the conserved charge number of the $J$-th conserved current.
$\phi (f^i) = f^i\ln f^i - \epsilon^{-1}
(1 + \epsilon f^i) \ln (1 + \epsilon f^i)$
where the sign factor $\epsilon$ is $+1$ for bosons, $-1$
for fermions and $0$ for classical particles.
In the classical limit, $\phi (f^i)$ reduces to
$f^i\ln f^i - f^i$.

We write the expansion of the entropy current up to the second order in 
$\delta f^i = f^i - f_0^i$ as
\begin{eqnarray}
\label{eq:ent_exp}
s^\mu &=& s_0^\mu + \delta s^\mu_{(1)} + \delta s^\mu_{(2)} + \mathcal{O}(\delta f^3) .
\end{eqnarray}
If one naively considers the first order dissipative correction then one obtains 
the Navier-Stokes expressions of the relativistic constitutive equations 
from the law of increasing entropy. Here we would like to emphasize that
as is the case for standard statistical mechanics, one needs to have 
Onsager cross terms in the first order expressions, even though 
they are sometimes neglected in conventional relativistic formalisms.
The cross terms are actually essential because they give rise to important physics such as 
Soret and Dufour effects and also play a significant role in preserving the law of 
increasing entropy. Detailed discussion on the relativistic linear response theory 
can be found in Appendix~\ref{sec:phenom}. However, the formalism is known 
to be both acausal and unstable \cite{HiscockLindblom}
as they allow propagation of information faster than the speed of light.
To avoid such unnatural behavior,
one has to consider the second order correction
to $s^\mu$ which introduces relaxation effects on the dissipative currents.

Israel-Stewart formalism of the second order
single-component dissipative hydrodynamics 
can be derived from the third moment of
the distribution $f^i$.
We generalize the formalism to multi-component systems and write
\begin{eqnarray}
\label{eq:third_multi}
\partial _\alpha I^{\mu \nu \alpha} = \sum _ i \int \frac{g_i d^3 p}{(2\pi )^3 E_i} p_i^\mu p_i^\nu p_i^\alpha \partial _\alpha f^i = Y^{\mu \nu}.
\end{eqnarray}
Here $Y^{\mu \nu}$ is a symmetric tensor
which we will determine later.
There are 10 independent equations,
which are, when projected either parallel/perpendicular to
the flow $u^\mu$,
the scalar equations $u_\mu u_\nu \partial _\alpha I^{\mu \nu \alpha}
= u_\mu u_\nu Y^{\mu \nu}$ (1 equation)
and $\Delta _{\mu \nu} \partial _\alpha I^{\mu \nu \alpha}
= \Delta_{\mu \nu} Y^{\mu \nu}$ (1 equation),
the vector equation $\Delta_{\rho \mu} u_\nu \partial _\alpha
I^{\mu \nu \alpha} = \Delta_{\rho \mu} u_\nu Y^{\mu \nu}$ (3 equations)
and the tensor equation 
$\partial _\alpha I^{\langle \mu \nu \rangle \alpha}
= Y^{\langle \mu \nu \rangle}$ (5 equations).

On the other hand, the number of unknown dissipative currents is $10+4N$
in the systems with non-vanishing chemical potentials.
In this respect, we need $4N$ more equations to
fully determine the system.
Since the conventional moment equations are
the derivatives of the higher order moment 
for the energy-momentum conservation 
as shown in Eq.~(\ref{eq:third_multi}),
it would be natural to introduce new moment
equations in multi-component systems,
\begin{eqnarray}
\label{eq:third_multi_b}
\partial _\alpha I^{\mu \alpha}_J 
&=& \sum _ i \int \frac{q_i^J g_i d^3 p}
{(2\pi )^3 E_i} p_i^\mu p_i^\alpha \partial _\alpha f^i = Y_J^{\mu},
\end{eqnarray}
which are the derivatives of the higher order moment
for the charge number conservations.
One might argue that weight factors other than
the conserved charge number $q_i^J$ could have been chosen
to construct the moment equations,
but Eq.~(\ref{eq:third_multi_b}) is
the only vector equation which vanishes
in the limit of zero net charge number densities.
This formalism not only makes
the correspondence between the number of equations
and that of unknowns clear,
but also allows one to construct an arbitrary number of
moment equations of the form 
$\partial_\alpha I^{\mu \alpha}_J = Y_J^\mu$
depending on how many conserved charge currents the system possesses.

We now have $10+4N$ equations for $10+4N$ dissipative currents.
The next task is to estimate $Y^{\mu \nu}$ and $Y_J^\mu$
to derive the constitutive equations
in terms of the dissipative currents.
The constraints are again given by the second law of thermodynamics.
The definition of the entropy current (\ref{eq:entropycurrent}) gives
\begin{eqnarray}
\label{eq:entropy}
\partial _\mu s^\mu &=& - \sum _i \int \frac{g_i d^3 p_i}{(2\pi )^3 E_i}
p_i^\mu \frac{\partial \phi}{\partial f^i} \partial_\mu f^i \nonumber \\
&=& \sum _i \int \frac{g_i d^3 p_i}{(2\pi )^3 E_i} p_i^\mu y^i
\partial_\mu f^i ,
\end{eqnarray}
where $y^i$ is defined in $f^i = [\exp (y^i) - \epsilon ]^{-1}$.
As discussed in Appendix~\ref{sec:phenom}, the second order constitutive equations depend on
an explicit form of the off-equilibrium distribution $f^i$,
or equivalently, $y^i$. When the deviation from the local thermal equilibrium is small, we may write $y^i$ as
\begin{equation}
\label{eq:y}
y^i = y_0^i + \delta y^i,
\end{equation} 
where $y_0^i$ is defined in $f_0^i =[\exp{(y_0^i)} - \epsilon ]^{-1}$, which means 
\begin{eqnarray}
\label{eq:y0}
y_0^i &=& - \sum _J \frac{q_i^J \mu _J}{T} + p_i^\mu \frac{u_\mu}{T}.
\end{eqnarray} 
$\mu_J$ is the chemical potential associated with the $J$-th
conserved quantity.
We estimate the off-equilibrium correction $\delta y^i$ through Grad's moment method.
The conventional Grad's 14-moment method \cite{Israel:1979wp, deGroot}
cannot be generally applied to the systems with $N$ conserved
charge currents because the number of macroscopic variables
is then $10+4N$ whereas that of unknowns in the expansion remains 14.
Actually, if the system has no conserved charge currents,
\textit{i.e.}, $N=0$, the system also cannot be solved,
unless the concept of one conserved current is present in the system.
This means that the system of single component with
binary collisions or multi-components
with one conserved charge current 
is implicitly assumed in the
Grad's 14-moment method.
In the latter case, if
there is no conserved charge current, 
the limit of vanishing chemical potential
for the stability condition can be taken \cite{Monnai:2009ad}.
Thus we have to introduce $10+4N$ unknowns in the expansion as well.
If we demand that (i) the distortion of the distribution
can be expressed in terms of the dissipative currents,
(ii) the resulting constitutive equations for the dissipative
currents satisfy the Onsager reciprocal relations
and (iii) an arbitrary number of conserved currents
can be introduced to the system and the effects of a conserved current
vanish in the vanishing limit of the corresponding chemical potential,
then the only possible way is to assume the expansion
\begin{eqnarray}
\label{eq:dym}
\delta y^i &=& p_i^\mu \sum _J q_i^J \varepsilon _{\mu}^J + p_i^\mu p_i^\nu \varepsilon _{\mu \nu} ,
\end{eqnarray}
where $\varepsilon^J_{\mu}$ and $\varepsilon_{\mu \nu}$ are
macroscopic coefficients of the expansion which include
information of all the components in the system.
We will see later that this indeed yields
the second order constitutive equations
which are reasonable compared with those of other formalisms. 

We make several comments on this distortion $\delta f^i$.
Firstly, a non-zero trace tensor correction $\varepsilon_{\mu \nu}$
is considered instead of a scalar and traceless tensor correction in Eq.~(\ref{eq:dym}).
We will see later that this must be the case
for multi-component systems \cite{Monnai:2009ad}
because of the law of increasing entropy.
Secondly, if one considers a single component system
with binary collisions, \textit{i.e.}, $q=1$,
then Eq.~(\ref{eq:dym}) reduces to the expansion
$\delta y = p^\mu \varepsilon _{\mu} + p^\mu p^\nu \varepsilon _{\mu \nu}$
which is equivalent to the conventional Grad's 14-moment method mentioned
in Ref.~\cite{Israel:1979wp}.
Note that this is not completely equivalent to the Grad's 14-moment method
for the systems with chemical interaction.
Thirdly, the distribution function satisfies the stability conditions
because $\delta e$ and $\delta n_J$ are treated carefully in this formalism
and they can be considered zero after the constitutive equations are derived.  

The unknowns in the distortion of the distribution
can be determined by matching the macroscopic variables
and the ones calculated in relativistic kinetic theory.
We first tensor-decompose the $10+4N$ unknowns $\varepsilon _{\mu \nu}$ and $\varepsilon^J _{\mu}$
in terms of the flow as
\begin{eqnarray}
\label{eq:quadraticepsilon}
\varepsilon _{\mu \nu} &=& \varepsilon^{\parallel \parallel} u_\mu u_\nu
+ \varepsilon ^{\perp \top} \frac{\Delta_{\mu \nu}}{3}
+ 2\varepsilon _{( \mu} ^{\perp \parallel} u_{\nu)}
+ \varepsilon_{\mu \nu}^{\perp \perp} , \\
\label{eq:linearepsilon}
\varepsilon^J _{\mu} &=& \varepsilon_J ^{\parallel} u_\mu
+ \varepsilon _{J \mu} ^\perp ,
\end{eqnarray}
where we employed the notations $\varepsilon _{\perp \top}
=  \Delta^{\mu \nu} \varepsilon _{\mu \nu}$,
$\varepsilon _{\parallel \parallel} = u^\mu u^\nu \varepsilon _{\mu \nu}$,
$\varepsilon^J _\parallel = u^\mu \varepsilon^J _\mu$,
$\varepsilon ^\mu _{\perp \parallel} = \Delta ^{\mu \nu} u^\rho
\varepsilon _{\nu \rho}$,
$\varepsilon ^{J \mu} _{\perp} = \Delta ^{\mu \nu} \varepsilon^J _\nu$
and $\varepsilon_{\perp \perp} ^{\mu \nu}
= \varepsilon ^{\langle \mu \nu \rangle}$.
Then the consistency conditions can be expressed as
\begin{eqnarray}
\label{eq:matricesm0}
\mathcal{J}_0
\left(
 \begin{array}{c}
  \varepsilon _{\perp \top} \\
  \varepsilon _{\parallel \parallel} \\
  \varepsilon _\parallel ^{J_1} \\
  \varepsilon _\parallel ^{J_2} \\
  \cdots
 \end{array}
\right)
&=&
\left(
 \begin{array}{c}
  - \Pi \\
  \delta e \\
  \delta n_{J_1} \\
  \delta n_{J_2} \\
  \cdots
 \end{array}
\right) , \\
\label{eq:matricesm1}
\mathcal{J}_1
\left(
 \begin{array}{c}
  \varepsilon ^\mu _{\perp \parallel} \\
  \varepsilon ^{J_1 \mu} _{\perp} \\
  \varepsilon ^{J_2 \mu} _{\perp} \\
  \cdots
 \end{array}
\right)
 &=&
\left(
 \begin{array}{c}
  W^\mu \\
  V_{J_1}^\mu \\
  V_{J_2}^\mu \\
  \cdots
 \end{array}
\right) ,
\\
\label{eq:matricesm2}
 \mathcal{J}_2 \varepsilon_{\perp \perp} ^{\mu \nu} 
 &=&
 \pi^{\mu \nu},
\end{eqnarray}
where the matching matrices are defined as 
\begin{eqnarray}
\label{eq:matricesm0d}
\mathcal{J}_0
&=&
- \left(
 \begin{array}{ccccc}
  \frac{5}{3}J_{42} & J_{41} & J^{j_1}_{31} & J^{j_2}_{31} & \cdots \\
  J_{41} & J_{40} & J^{j_1}_{30} & J^{j_2}_{30} & \cdots \\
  {J}^{j_1}_{31} & {J}^{j_1}_{30} & {J}^{j_1 j_1}_{20} & {J}^{j_1 j_2}_{20} & \cdots \\
  {J}^{j_2}_{31} & {J}^{j_2}_{30} & {J}^{j_1 j_2}_{20} & {J}^{j_2 j_2}_{20} & \cdots \\
  \cdots & \cdots & \cdots & \cdots & \cdots 
 \end{array}
\right) ,
\\
\label{eq:matricesm1d}
\mathcal{J}_1
 &=&
- \left(
 \begin{array}{cccc}
  2J_{41} & J^{j_1}_{31} & J^{j_2}_{31} & \cdots \\
  2J^{j_1}_{31} & J^{j_1 j_1}_{21} & J^{j_1 j_2}_{21} & \cdots \\
  2J^{j_2}_{31} & J^{j_1 j_2}_{21} & J^{j_2 j_2}_{21} & \cdots \\
  \cdots & \cdots & \cdots & \cdots
 \end{array}
\right) ,
\\
\label{eq:matricesm2d}
\mathcal{J}_2
 &=&
 -2J_{42} .
\end{eqnarray}
Here the moments of the distribution are defined as 
\begin{eqnarray}
\label{eq:jmn}
J_{jk \cdots}^{\mu _1 \cdots \mu _m} &= & \sum _i 
\int \frac{(q^{J}_i q^{K}_i \cdots)g_i d^3 p}{(2\pi )^3 E_i}
 f_0^i (1+\epsilon f_0^i) p_i^{\mu _1} \cdots p_i^{\mu _m} \nonumber \\
&= & \sum _n \big[ (\Delta ^{\mu _1 \mu _2} \cdots 
\Delta ^{\mu _{2n-1} \mu _{2n}} u^{\mu _{2n+1}} \cdots u^{\mu _m}) \nonumber \\
&+ & \mathrm{(permutations)} \big] J_{mn}^{jk \cdots},
\end{eqnarray}
where the index $jk\cdots$ denotes 
the additional weight factor $q^{J}_i q^{K}_i \cdots$
in the summation over particle species $i$.
The moments with no index mean no weight factor in the summation.
Note that the moments with double charge weight,
\textit{e.g.}, $q^{J}_i q^{J}_i$ do not vanish 
in the limit of vanishing chemical potential. 

Then Eqs.~(\ref{eq:matricesm0})-(\ref{eq:matricesm2}) 
can be easily solved and we obtain 
\begin{eqnarray}
\varepsilon _{\mu \nu} &=& (B_\Pi \Pi + B_{\delta e} 
\delta e + \sum _J B_{\delta n_J} \delta n_J ) \Delta _{\mu \nu} \nonumber \\
\label{eq:quadratic}
&+& (\tilde{B}_\Pi \Pi + \tilde{B}_{\delta e} \delta e
 + \sum _J \tilde{B}_{\delta n_J} \delta n_J ) u_\mu u_\nu \nonumber \\
&+& 2 B_W u_{( \mu} W_{\nu )} + 2 \sum_J B_{V_J} u_{( \mu} V^J_{\nu )}
 + B_\pi \pi_{\mu \nu}, \\
\label{eq:linear}
\varepsilon^J _{\mu} &=& (D^J_\Pi \Pi + D^J_{\delta e} \delta e
 + \sum _{K} D^J_{\delta n_{K}} \delta n_{K} ) u_\mu \nonumber \\
&+& D^J_W W_\mu + \sum_{K} D^J_{V_{K}} V^{K}_\mu ,
\end{eqnarray}
where $B$'s and $D$'s are macroscopic quantities
which can be determined by temperature and
chemical potential only and are momentum independent.
They contain information of all the constituent particles in the system.
The explicit expressions of the prefactors
for the dissipative currents in relativistic kinetic theory are 
\begin{eqnarray}
B_{\Pi} &=& -\frac{1}{3} (\mathcal{J}_0^{-1})_{11}, \ \ B_{\delta e} 
=  \frac{1}{3} (\mathcal{J}_0^{-1})_{12}, \\
B_{\delta n_J} &=&  \frac{1}{3} (\mathcal{J}_0^{-1})_{1,2+j}, \ \
  \tilde{B}_{\Pi} = - (\mathcal{J}_0^{-1})_{21}, \\
\tilde{B}_{\delta e} &=&  (\mathcal{J}_0^{-1})_{22}, \ \
\tilde{B}_{\delta n_J} = (\mathcal{J}_0^{-1})_{2,2+j}, \\
B_W &=& (\mathcal{J}_1^{-1})_{11}, \ \ B_{V_J} 
= (\mathcal{J}_1^{-1})_{1,1+j}, \\ 
B_{\pi} &=& (\mathcal{J}_2^{-1})_{11},
\ \ D^J_{\Pi} = - (\mathcal{J}_0^{-1})_{2+j,1},\\
D^J_{\delta e} &=& (\mathcal{J}_0^{-1})_{2+j,2},
\ \ D^J_{\delta n_{K}} = (\mathcal{J}_0^{-1})_{2+j,2+k}, \\
D^J_W &=& (\mathcal{J}_1^{-1})_{1+j,1},
\ \ D^J_{V_K} = (\mathcal{J}_1^{-1})_{1+j,1+k},
\end{eqnarray}
where $1\leq (k, l) \leq N$.
It should be noted that $\delta y^i$ is expressed as
linear combinations of the dissipative currents. 
Higher order contributions are discussed in Appendix~\ref{sodod}.

The entropy production (\ref{eq:entropy}) is now expressed as
\begin{eqnarray}
\label{eq:entropy2}
\partial _\mu s^\mu 
& = & \sum _i \int \frac{g_i d^3p}{(2\pi )^3 E_i} (y_{0}^{i} + \delta y^{i})
p_{i}^{\mu} \partial_{\mu} f^{i} \nonumber \\
&=& \sum _i \int \frac{g_i d^3p}{(2\pi )^3 E_i}
\bigg[ - \bigg( \sum_J \frac{q_i^J \mu_J}{T} \bigg) \nonumber \\
&+& \bigg( \frac{u_\nu}{T} + \sum_J q_i^J \varepsilon^J_\nu \bigg)
p_i^\nu + \varepsilon_{\nu \rho} p_i^\nu p_i^\rho \bigg]
p_i^\mu \partial_\mu f^i \nonumber \\
&=& \sum_J \varepsilon^J_\nu Y_J^\nu 
+ \varepsilon_{\nu \rho} Y^{\nu \rho} \geq 0,
\end{eqnarray}
where we have used the energy-momentum conservation
and the charge number conservations.
The tensor structures of $\varepsilon_{\nu \rho}$
and $\varepsilon_\nu^J$ in terms of the flow $u^\mu$
needs to be reflected on $Y^{\nu \rho}$ and $Y_J^\nu$
in linear response theory.
Note that again the ``cross terms" are allowed
for the reason mentioned in Appendix~\ref{sec:phenom}.
The finite trace tensor correction to the distribution
$\varepsilon^{\mu \nu}$ is required
instead of the traceless one with the scalar correction $\varepsilon$
for systems with multi-components and/or multiple conserved currents,
because $Y^{\mu}_{\mu} \neq 0$ can be shown from
the fact that the trace of Eq.~(\ref{eq:third_multi})
is not generally zero, \textit{i.e.},
\begin{equation}
\label{eq:trace}
\partial _\alpha I^{\mu \alpha}_\mu 
= \sum _ i m_i ^2 \int \frac{g_i d^3 p}{(2\pi )^3 E_i}
p_i^\alpha \partial _\alpha f^i \neq 0 ,
\end{equation}
and the existence of the moment equation for $\varepsilon$
violates the matching of the number of dissipative currents
and that of the constitutive equations.
It gives a justification to the expansion Eq.~(\ref{eq:dym})
that a non-zero trace $\varepsilon^{\mu \nu}$ should be considered. 
The tensor structures of $Y^{\mu \nu}$ and $Y_J^\mu$
are then expressed as
\begin{eqnarray}
\label{eq:matricesc0}
\left(
 \begin{array}{c}
  Y _{\perp \top} \\
  Y _{\parallel \parallel} \\
  Y _\parallel ^{J_1} \\
  Y _\parallel ^{J_2} \\
  \cdots
 \end{array}
\right)
&=&
\mathcal{C}_0^{-1}
\left(
 \begin{array}{c}
  \varepsilon _{\perp \top} \\
  \varepsilon _{\parallel \parallel} \\
  \varepsilon _\parallel ^{J_1} \\
  \varepsilon _\parallel ^{J_2} \\
  \cdots
 \end{array}
\right) , \\
\label{eq:matricesc1}
\left(
 \begin{array}{c}
  Y ^\mu _{\parallel \perp} \\
  Y ^{J_1 \mu} _{\perp} \\
  Y ^{J_2 \mu} _{\perp} \\
  \cdots
 \end{array}
\right)
 &=&
-\mathcal{C}_1^{-1}
\left(
 \begin{array}{c}
  \varepsilon ^\mu _{\parallel \perp} \\
  \varepsilon ^{J_1 \mu} _{\perp} \\
  \varepsilon ^{J_2 \mu} _{\perp} \\
  \cdots
 \end{array}
\right) ,
\\
\label{eq:matricesc2}
 Y_{\perp \perp} ^{\mu \nu} 
 &=&
 \mathcal{C}_2^{-1}
 \varepsilon_{\perp \perp} ^{\mu \nu} ,
\end{eqnarray}
where 
\begin{eqnarray}
\label{eq:quadraticY}
Y _{\mu \nu} &=& Y^{\parallel \parallel} u_\mu u_\nu
+ Y ^{\perp \top} \frac{\Delta_{\mu \nu}}{3}
+ 2Y _{( \mu} ^{\parallel \perp} u_{\nu)}
+ Y_{\mu \nu}^{\perp \perp} , \\
\label{eq:linearY}
Y^J _{\mu} &=& Y_J ^{\parallel} u_\mu + Y _{J \mu} ^\perp .
\end{eqnarray}
The microscopic physics is integrated out 
in the transport coefficient matrices $\mathcal{C}_i$'s.
Here $\mathcal{C}_i$'s are semi-positive definite
and symmetric because of Onsager reciprocal relations.
$Y^\mu_J \to 0$ and $\varepsilon_\mu^J \to 0$
in the limit of corresponding vanishing chemical potential $\mu_J$
implies that the transport coefficients
for the cross terms between $\varepsilon_{\mu \nu}$ and $Y^J _{\mu}$, 
or equivalently, $\varepsilon^J _{\mu}$ and $Y_{\mu \nu}$, 
also vanish in the limit.
We uniquely determine the constitutive equations
for all the dissipative currents 
by solving Eqs.~(\ref{eq:matricesc0})-(\ref{eq:matricesc2})
in terms of the $10+4N$ dissipative currents
$\varepsilon^{\mu \nu}$ and $\varepsilon_J^\mu$
and then in terms of $\Pi$, $\delta e$, $W^\mu$,
$\pi^{\mu \nu}$, $\delta n_J$ and $V_J^\mu$
using the explicit forms of the distortion of
the phase-space distribution (\ref{eq:quadratic})-(\ref{eq:linear}). 

We obtain the constitutive equations explicitly by estimating 
the derivatives $\partial _\alpha I^{\mu \nu \alpha}$ and $\partial _\alpha I_J^{\mu \alpha}$
in Eqs.~(\ref{eq:third_multi}) and (\ref{eq:third_multi_b}).
Remembering $y^i = - \sum_J \frac{q_i^J \mu _J}{T} 
+ p_i^\mu ( \frac{u_\mu}{T} + \sum_J q_i^J \varepsilon^J _{\mu} ) 
+ p_i^\mu p_i^\nu \varepsilon _{\mu \nu}$,
their expressions up to the second order are
\begin{eqnarray}
\label{eq:lhs}
\partial _\alpha I^{\mu \nu \alpha} 
&=& \sum_J J_j^{\mu \nu \alpha} \partial _\alpha \frac{\mu _J}{T}
- J^{\mu \nu \alpha \beta} \partial _\alpha \frac{u _\beta}{T} \nonumber \\
&-& \sum_J J_j^{\mu \nu \alpha \beta} \partial _\alpha \varepsilon^J _\beta
- J^{\mu \nu \alpha \beta \gamma } 
\partial _\alpha \varepsilon _{\beta \gamma} \nonumber \\
&-& \bigg( \sum_{J,K} K_{jk}^{\mu \nu \alpha \beta} 
\varepsilon^{J} _\beta + \sum_{K} K_{k} ^{\mu \nu \alpha \beta \gamma}
\varepsilon _{\beta \gamma} \bigg) \partial _\alpha \frac{\mu _{K}}{T} \nonumber \\
&+& \bigg( \sum_J K_j^{\mu \nu \alpha \beta \gamma} 
\varepsilon^J _\gamma + K^{\mu \nu \alpha \beta \gamma \delta}
\varepsilon_{\gamma \delta} \bigg) \partial _\alpha \frac{u _\beta}{T} ,
\end{eqnarray}
\begin{eqnarray}
\label{eq:lhs_2}
\partial _\alpha I_J^{\mu \alpha} 
&=& \sum_{K} J_{jk}^{\mu \alpha}
\partial _\alpha \frac{\mu _{K}}{T} - J_j^{\mu \alpha \beta}
\partial _\alpha \frac{u _\beta}{T} \nonumber \\
&-& \sum_{K} J_{jk}^{\mu \alpha \beta} 
\partial _\alpha \varepsilon^{K} _\beta - J_j^{\mu \alpha \beta \gamma }
\partial _\alpha \varepsilon _{\beta \gamma} \nonumber \\
&-& \bigg( \sum_{K,L} K_{jkl}^{\mu \alpha \beta}
\varepsilon^{K} _\beta + \sum_{L} K_{jl} ^{\mu \alpha \beta \gamma}
\varepsilon _{\beta \gamma} \bigg) \partial _\alpha
\frac{\mu _{L}}{T} \nonumber \\
&+& \bigg( \sum_{K} K_{jk}^{\mu \alpha \beta \gamma}
\varepsilon^{K} _\gamma + K_j^{\mu \alpha \beta \gamma \delta}
\varepsilon_{\gamma \delta} \bigg) \partial _\alpha \frac{u _\beta}{T} ,
\end{eqnarray}
where the additional moments are defined as
\begin{eqnarray}
\label{eq:kmn}
K_{jk\cdots}^{\mu _1 \cdots \mu _m} 
&= & \sum _i \int \frac{(q^{J}_i q^{K}_i \cdots) g_i d^3 p}{(2\pi )^3 E_i}
f_0^i (1+\epsilon f_0^i) \nonumber \\
&\times & (1+ 2\epsilon f_0^i) p_i^{\mu _1} \cdots p_i^{\mu _m} \nonumber \\
&= & \sum _n \big[ (\Delta ^{\mu _1 \mu _2} \cdots 
\Delta ^{\mu _{2n-1} \mu _{2n}} u^{\mu _{2n+1}} ... u^{\mu _m}) \nonumber \\
&+ & \mathrm{(permutations)} \big] K^{jk\cdots}_{mn}.
\end{eqnarray}
Again the index $jk\cdots$ denotes the additional weight factors
$q^{J}_i q^{K}_i \cdots$ in the summation over particle species.
The terms in the third and fourth lines of Eq.~(\ref{eq:lhs})
which involve the moments $K$'s
are assumed to be small and simply omitted
in the Israel-Stewart formalism \cite{Israel:1979wp}.
It is argued in that paper that the terms proportional to 
the Navier-Stokes thermodynamic forces, \textit{i.e.}, 
$D\frac{1}{T}$, $D\frac{\mu}{T}$, $\nabla_\mu u^\mu$, $\nabla_\mu \frac{1}{T}$, 
$\nabla_\mu \frac{\mu}{T}$ and $\nabla_{\langle \mu} u_{\nu \rangle}$ 
may not be significant, even though the terms proportional to acceleration 
$D u^\mu$ are kept in their equations.
However, these are actually of the same order in dissipative currents
as the other second-order terms 
and should not be neglected to preserve consistency.
Equation (\ref{eq:lhs_2}) corresponds to
the new moment equations
which do not appear in the original Israel-Stewart theory.

It is now a straight-forward task
to derive the constitutive equations
in multi-component systems with multiple conserved currents.
We have from Eqs.~(\ref{eq:third_multi}), (\ref{eq:third_multi_b}),
(\ref{eq:matricesm0})-(\ref{eq:matricesm2}) and
(\ref{eq:matricesc0})-(\ref{eq:matricesc2}),
\begin{eqnarray}
\label{eq:matrices0}
\left(
 \begin{array}{c}
  - \Pi \\
  \delta e \\
  \delta n_1 \\
  \delta n_2 \\
  \cdots
 \end{array}
\right) &=&
\mathcal{J}_{0} \mathcal{C}_{0}
\left(
 \begin{array}{c}
  \frac{1}{3} \partial _\alpha I^{\perp \top \alpha} \\
  \partial _\alpha I^{\parallel \parallel \alpha} \\
  \partial _\alpha I_{J_1}^{\parallel \alpha} \\
  \partial _\alpha I_{J_2}^{\parallel \alpha} \\
  \cdots
 \end{array}
\right) ,
\\
\label{eq:matrices1}
\left(
 \begin{array}{c}
  W^\mu \\
  V_1^\mu \\
  V_2^\mu \\
  \cdots
 \end{array}
\right) 
 &=&
- \mathcal{J}_{1} \mathcal{C}_{1}
\left(
 \begin{array}{c}
  2 \partial _\alpha I^{\perp \parallel \mu \alpha} \\
  \partial _\alpha I_{J_1}^{\perp \mu \alpha} \\
  \partial _\alpha I_{J_2}^{\perp \mu \alpha} \\
  \cdots
 \end{array}
\right)  ,
\\
\label{eq:matrices2}
 \pi^{\mu \nu}
 &=&
 \mathcal{J}_2 \mathcal{C}_2 \partial _\alpha I^{\perp \perp \mu \nu \alpha},
\end{eqnarray}
and when combined with Eqs. (\ref{eq:lhs}) and (\ref{eq:lhs_2}),
the second-order constitutive equations 
for the dissipative currents are expressed as follows:
\begin{eqnarray}
\Pi &=& -\zeta \nabla _\mu u^\mu - \tau_\Pi D \Pi \nonumber \\
&+& \sum_J \chi_{\Pi \Pi}^{aJ} \Pi D\frac{\mu _J}{T} + \chi_{\Pi \Pi}^b \Pi D \frac{1}{T} + \chi_{\Pi \Pi}^c \Pi \nabla _\mu u^\mu \nonumber \\
&-& \zeta_{\Pi \delta e} D\frac{1}{T} + \sum _J \zeta_{\Pi \delta n_J} D\frac{\mu_J}{T} \nonumber \\
&+& \sum_J \chi_{\Pi W}^{aJ} W_\mu \nabla ^\mu \frac{\mu _J}{T} + \chi_{\Pi W}^b W_\mu \nabla ^\mu \frac{1}{T} \nonumber \\
&+& \chi_{\Pi W}^c W_\mu D u ^\mu + \chi_{\Pi W}^d \nabla ^\mu W_\mu \nonumber \\
&+& \sum_{J,K} \chi_{\Pi V_J}^{aK} V^J_\mu \nabla ^\mu \frac{\mu_{K}}{T} + \sum_J \chi_{\Pi V_J}^b V^J_\mu \nabla ^\mu \frac{1}{T} \nonumber \\
&+& \sum_J \chi_{\Pi V_J}^c V^J_\mu D u ^\mu + \sum_J \chi_{\Pi V_J}^d \nabla ^\mu V^J_\mu \nonumber \\
&+& \chi_{\Pi \pi} \pi _{\mu \nu} \nabla ^{\langle \mu} u^{\nu \rangle} ,
\label{eq:Pi}
\end{eqnarray}
\begin{eqnarray}
W^\mu &=& -\kappa _W \bigg( \frac{1}{T} D u^\mu + \nabla ^\mu \frac{1}{T} \bigg) - \tau_{W} \Delta^{\mu \nu} D W_\nu \nonumber \\
&+& \sum_J \chi_{W W}^{aJ} W^\mu D\frac{\mu _J}{T} \nonumber \\
&+& \chi_{WW}^b W^\mu D \frac{1}{T} + \chi_{W W}^c W^\mu \nabla _\nu u^\nu \nonumber \\
&+& \chi_{WW}^d W^\nu \nabla _\nu  u^\mu + \chi_{WW}^e W^\nu \nabla ^\mu u_\nu \nonumber \\
&+& \sum_J \kappa_{W V_J} \nabla ^\mu \frac{\mu _J}{T} - \sum_J \tau _{W V_J} \Delta^{\mu \nu} D V^J_\nu \nonumber \\
&+& \sum_{J,K} \chi_{WV_{J}}^{aK} V_J^\mu D\frac{\mu _{K}}{T} \nonumber \\
&+& \sum_J \chi_{WV_J}^b V_J^\mu D \frac{1}{T} + \sum_J \chi_{W V_J}^c V_J^\mu \nabla ^\nu u_\nu \nonumber \\
&+& \sum_J \chi_{WV_J}^d V_J^\nu \nabla _\nu  u^\mu + \sum_J \chi_{WV_J}^e V_J^\nu \nabla ^\mu u_\nu \nonumber \\
&+& \sum_J \chi_{W \pi}^{aJ} \pi^{\mu \nu} \nabla_\nu \frac{\mu _J}{T} + \chi_{W \pi}^b \pi^{\mu \nu} \nabla_\nu \frac{1}{T} \nonumber \\ 
&+& \chi_{W \pi}^c \pi ^{\mu \nu} D u_\nu + \chi_{W \pi}^d \Delta^{\mu \nu} \nabla ^\rho \pi _{\nu \rho} \nonumber \\
&+& \sum_J \chi_{W \Pi}^{aJ} \Pi \nabla ^\mu \frac{\mu _J}{T} + \chi_{W \Pi}^b \Pi \nabla ^\mu \frac{1}{T} \nonumber \\
&+& \chi_{W \Pi}^c \Pi D u^\mu + \chi_{W \Pi}^d \nabla ^\mu \Pi ,
\label{eq:W}
\end{eqnarray}
\begin{eqnarray}
V_J^\mu &=& \kappa_{V_J} \nabla ^\mu \frac{\mu _J}{T} - \tau _{V_J} \Delta^{\mu \nu} D V^{J}_\nu \nonumber \\
&+& \sum_{K\neq J} \kappa_{V_J V_K} \nabla ^\mu \frac{\mu _K}{T} - \sum_{K\neq J} \tau _{V_J V_{K}} \Delta^{\mu \nu} D V^{K}_\nu \nonumber \\
&+& \sum_{K,L} \chi_{V_J V_{K}}^{aL} V_{K}^\mu D \frac{\mu _{L}}{T} \nonumber \\
&+& \sum_{K} \chi_{V_J V_{K}}^b V_{K}^\mu D \frac{1}{T} + \sum_{K} \chi_{V_J V_{K}}^c V_{K}^\mu \nabla _\nu u^\nu \nonumber \\
&+& \sum_{K} \chi_{V_J V_{K}}^d V_{K}^\nu \nabla _\nu u^\mu + \sum_{K} \chi_{V_J V_{K}}^e V_{K}^\nu \nabla ^\mu u_\nu \nonumber \\
&-& \kappa _{V_J W} \bigg( \frac{1}{T} D u^\mu + \nabla ^\mu \frac{1}{T} \bigg) - \tau_{V_J W} \Delta^{\mu \nu} D W_\nu \nonumber \\
&+& \sum_{K} \chi_{V_J W}^a W^\mu D \frac{\mu _{K}}{T} \nonumber \\
&+& \chi_{V_J W}^b W^\mu D \frac{1}{T} + \chi_{V_J W}^c W^\mu \nabla ^\nu u_\nu \nonumber \\
&+& \chi_{V_J W}^d W^\nu \nabla _\nu u^\mu + \chi_{V_J W}^e W^\nu \nabla ^\mu u_\nu \nonumber \\
&+& \sum_{K} \chi_{V_J \pi}^{aK} \pi^{\mu \nu} \nabla_\nu \frac{\mu _{K}}{T} + \chi_{V_J \pi}^b \pi^{\mu \nu} \nabla_\nu \frac{1}{T} \nonumber \\
&+& \chi_{V_J \pi}^c \pi ^{\mu \nu} D u_\nu \nonumber + \chi_{V_J \pi}^d \Delta^{\mu \nu} \nabla ^\rho \pi _{\nu \rho} \\
&+& \sum_{K} \chi_{V_J \Pi}^{aK} \Pi \nabla ^\mu \frac{\mu _{K}}{T} + \chi_{V_J \Pi}^b \Pi \nabla ^\mu \frac{1}{T} \nonumber \\
&+& \chi_{V_J \Pi}^c \Pi D u^\mu + \chi_{V_J \Pi}^d \nabla ^\mu \Pi ,
\label{eq:V}
\end{eqnarray}
\begin{eqnarray}
\pi^{\mu \nu} &=& 2 \eta \nabla ^{\langle \mu} u^{\nu \rangle} - \tau_\pi D \pi^{\langle \mu \nu \rangle} \nonumber \\
&+& \sum_J \chi_{\pi \pi}^{aJ} \pi^{\mu \nu} D \frac{\mu_J}{T} + \chi_{\pi \pi}^b \pi^{\mu \nu} D \frac{1}{T} \nonumber \\
&+& \chi_{\pi \pi}^c \pi ^{\mu \nu} \nabla _\rho u^\rho + \chi_{\pi \pi}^d \pi ^{\rho \langle \mu} \nabla _\rho u^{\nu \rangle} \nonumber \\
&+& \sum_J \chi_{\pi W}^{aJ} W^{\langle \mu} \nabla ^{\nu \rangle} \frac{\mu_J}{T} + \chi_{\pi W}^b W^{\langle \mu} \nabla ^{\nu \rangle} \frac{1}{T}  \nonumber \\
&+& \chi_{\pi W}^c W^{\langle \mu} D u ^{\nu \rangle} + \chi_{\pi W}^d \nabla ^{\langle \mu} W^{\nu \rangle} \nonumber \\
&+& \sum_{J,K} \chi_{\pi V_J}^{aJ} V_J^{\langle \mu} \nabla ^{\nu \rangle} \frac{\mu_{K}}{T} + \sum_J \chi_{\pi V_J}^b V_J^{\langle \mu} \nabla ^{\nu \rangle} \frac{1}{T} \nonumber \\
&+& \sum_J \chi_{\pi V_J}^c V_J^{\langle \mu} D u^{\nu \rangle} + \sum_J \chi_{\pi V_J}^d \nabla ^{\langle \mu} V_J ^{\nu \rangle} \nonumber \\
&+& \chi _{\pi \Pi} \Pi \nabla ^{\langle \mu} u^{\nu \rangle} .
\label{eq:pi}
\end{eqnarray}
Here $\zeta$'s, $\kappa$'s and $\eta$ are the first order
transport coefficients which are expressed
in terms of the matching matrices $\mathcal{J}_i$ and 
the semi-positive definite matrices $\mathcal{C}_i$.
As we will see in Sec.~\ref{subsec:onsager}, these
transport coefficients satisfy Onsager reciprocal relations.
$\zeta$ is called bulk viscosity,
$\kappa _W$ energy conductivity,
$\eta$ shear viscosity
and $\kappa _{V_J}$ charge conductivity
of the $J$-th conserved current.
$\tau$'s are the relaxation times
and $\chi$'s are the second order transport coefficients. 
The stability conditions are employed at this point
to take out the constitutive equations for $\delta e$
and $\delta n_J$, and to obtain the second order
constitutive equations for $\Pi$, $W^\mu$, $V_J^\mu$ and $\pi^{\mu \nu}$.
Apparently we also have the term 
$\chi_{\pi \pi}^e \pi ^{\rho \langle \mu} \nabla^{\nu \rangle}  u_\rho$
in Eq.~(\ref{eq:pi}), but this term actually vanishes
because orthogonality relation $\pi_{\rho \mu} u^\rho = 0$
demands $\chi_{\pi \pi}^e=0$.
We have utilized the fact that the prefactors
$D$'s and $B$'s appearing in $\partial_\mu \delta f^i$
are functions of $J_{mn}^{kl \cdots}$'s,
and that their derivatives can be expressed
in terms of the Navier-Stokes thermodynamic forces:
\begin{eqnarray}
\label{Jmn_conv}
\partial _\mu F &=& \sum_{K,L,\cdots} \sum _{m,n} \frac{\delta F}{\delta {J}^{kl \cdots}_{mn}} \partial _\mu {J}^{kl \cdots}_{mn} \nonumber \\
&=& \sum_J \sum_{K,L,\cdots} \sum _{m,n} \frac{\delta F}{\delta J^{kl \cdots}_{mn}} K^{j kl \cdots}_{mn} \partial _\mu \frac{\mu _J}{T} \nonumber \\
&-& \sum_{K,L,\cdots} \sum _{m,n} \frac{\delta F}{\delta J^{kl \cdots}_{mn}} K^{kl \cdots}_{(m+1)n} \partial _\mu \frac{1}{T}.
\end{eqnarray}
Here $F$ denotes the prefactors $B$'s and $D$'s.
Note that if we define the symmetric traceless thermodynamic force
$\sigma ^{\mu \nu} = \frac{1}{2} 
(\nabla^\mu u^\nu + \nabla^\nu u^\mu)
-\frac{1}{3} \Delta^{\mu \nu} \nabla _\alpha u^\alpha$
and the vorticity $\omega ^{\mu \nu} = \frac{1}{2}
(\nabla^\mu u^\nu - \nabla^\nu u^\mu)$,
then $\pi ^{\rho \langle \mu} \nabla _\rho u^{\nu \rangle}$
in Eq.~(\ref{eq:pi}) is expressed as
\begin{equation}
\label{eq:tensor_vortisity_indentity}
\pi ^{\rho \langle \mu} \nabla _\rho u^{\nu \rangle}
= \pi ^{\rho \langle \mu} \sigma ^{\ \nu \rangle} _{\rho}
+ \pi ^{\rho \langle \mu} \omega ^{\ \nu \rangle} _{\rho}
+ \frac{1}{3} \pi^{\mu \nu} \nabla^\rho u_\rho ,
\end{equation}
which is also a commonly found expression.
Likewise, the identities
\begin{eqnarray}
\label{eq:vector_vortisity_indentity}
W^\nu \nabla ^\mu u_\nu
&=& W^\nu \sigma ^\mu _{\ \nu} + W^\nu \omega ^\mu _{\ \nu}
+ \frac{1}{3} W^\mu \nabla^\nu u_\nu , \\
\label{eq:vector_vortisity_indentity2}
V_J^\nu \nabla ^\mu u_\nu
&=& V_J^\nu \sigma ^\mu _{\ \nu} + V_J^\nu \omega ^\mu _{\ \nu}
+ \frac{1}{3} V_J^\mu \nabla^\nu u_\nu ,
\end{eqnarray}
are often used.
One should be careful that $\omega ^{\mu \nu}$ is an anti-symmetric tensor.
In the first order limit, these constitutive equations
reduce to the Navier-Stokes forms.

Our formalism have four major differences from the conventional method
by Israel and Stewart \cite{Israel:1979wp}.
Firstly, several second order terms which do not appear
in the Israel-Stewart theory \cite{Israel:1979wp}
can be found in Eqs.~(\ref{eq:Pi})-(\ref{eq:pi}).
Actually, these terms -- the terms composed of a dissipative current 
and a Navier-Stokes thermodynamic force --
are the results of consistent expansion
and should also exist in the case of single component systems.
We will compare our formalism with others in Sec.~\ref{sec3}
and see most of the second order terms reported in other papers
are found in our formalism.
Secondly, we can now calculate the independent second order equations
for the vector dissipative currents $W^\mu$ and $V_J^\mu$,
which allows us to determine these variables in an arbitrary frame,
whereas the Israel-Stewart method yields the three equations for $q^\mu$ only.
Thirdly, we have different kinetic expressions of
the transport coefficients $\zeta$'s, $\kappa$'s, 
$\eta$, $\tau$'s and $\chi$'s due to the new moment equations. 
Note that once the first order transport coefficients are given, 
one can estimate the second order ones and the relaxation times
because they are related within the framework of kinetic theory.
Fourthly, chemically interacting systems 
with multiple conserved currents can now be uniquely determined.

\subsection{Onsager Reciprocal Relations} 
\label{subsec:onsager}

We investigate the extended Israel-Stewart theory
in Sec.~\ref{subsec:eis} keeping terms up to the first order 
and see that Onsager reciprocal relations are indeed satisfied
in our formalism. 
The entropy production is, according to Eq.~(\ref{eq:entropy}),
expressed up to the first order as
\begin{eqnarray}
\partial _\mu s^\mu 
&=& \sum _i \int \frac{g_i d^3 p}{(2\pi )^3 E_i}
p_i^\mu \delta y^i \partial_\mu  f_0^i
+ \mathcal{O}[\partial (\delta f^2)] \nonumber \\
\label{eq:ent_pro_IS0}
&=& \sum_J \varepsilon^J_{\mu} \partial _\alpha
I_{J0}^{\mu \alpha} + \varepsilon_{\mu \nu}
\partial _\alpha I_0^{\mu \nu \alpha}.
\end{eqnarray}
Then semi-positive definiteness of the above equation yields
$10+4N$ constitutive equations.
It is straight forward to derive the first order constitutive
equations as
\begin{eqnarray}
\label{eq:matrices0}
\left(
 \begin{array}{c}
  - \Pi \\
  \delta e \\
  \delta n_1 \\
  \delta n_2 \\
  \cdots
 \end{array}
\right) &=&
\mathcal{A}_{0}
\left(
 \begin{array}{c}
  \frac{1}{T} \nabla _\mu u^\mu \\
  D \frac{1}{T} \\
  -D \frac{\mu _1}{T} \\
  -D \frac{\mu _2}{T} \\
  \cdots
 \end{array}
\right) ,
\\
\label{eq:matrices1}
\left(
 \begin{array}{c}
  W^\mu \\
  V_1^\mu \\
  V_2^\mu \\
  \cdots
 \end{array}
\right) 
 &=&
- \mathcal{A}_{1}
\left(
 \begin{array}{c}
  \nabla _\mu \frac{1}{T} + \frac{1}{T} D u_\mu \\
  -\nabla _\mu \frac{\mu _1}{T} \\
  -\nabla _\mu \frac{\mu _2}{T} \\
  \cdots
 \end{array}
\right) ,
\\
\label{eq:matrices2}
 \pi^{\mu \nu}
 &=&
 \mathcal{A}_2 \frac{1}{T} \nabla ^{\langle \mu} u^{\nu \rangle} .
\end{eqnarray}
The explicit forms of the transport coefficient matrices are expressed as 
\begin{eqnarray}
\mathcal{A}_0 &=& \mathcal{J}_0 \mathcal{C}_0 \mathcal{J}_0^{\bf T} , \\
\mathcal{A}_1 &=& \mathcal{J}_1 \mathcal{C}_1 \mathcal{J}_1^{\bf T} , \\
\mathcal{A}_2 &=& \mathcal{J}_2 \mathcal{C}_2 \mathcal{J}_2^{\bf T} , 
\end{eqnarray}
using the moment equations 
(\ref{eq:third_multi})-(\ref{eq:third_multi_b}),
the matching of dissipative currents with $\varepsilon$'s 
(\ref{eq:matricesm0})-(\ref{eq:matricesm2}),
the second law of thermodynamics
(\ref{eq:matricesc0})-(\ref{eq:matricesc2}) and 
\begin{eqnarray}
\label{eq:matricesj0}
\left(
 \begin{array}{c}
  \frac{1}{3} \partial _\alpha I_0^{\perp \top \alpha} \\
  \partial _\alpha I_0^{\parallel \parallel \alpha} \\
  \partial _\alpha I_{{J_1}0}^{\parallel \alpha} \\
  \partial _\alpha I_{{J_2}0}^{\parallel \alpha} \\
  \cdots
 \end{array}
\right) &=&
\mathcal{J}_0^{\bf T}
\left(
 \begin{array}{c}
  \frac{1}{T} \nabla _\mu u^\mu \\
  D \frac{1}{T} \\
  -D \frac{\mu _1}{T} \\
  -D \frac{\mu _2}{T} \\
  \cdots
 \end{array}
\right) ,
\\
\label{eq:matricesj1}
\left(
 \begin{array}{c}
  2 \partial _\alpha I_0^{\perp \parallel \mu \alpha} \\
  \partial _\alpha I_{{J_1}0}^{\perp \mu \alpha} \\
  \partial _\alpha I_{{J_2}0}^{\perp \mu \alpha} \\
  \cdots
 \end{array}
\right) 
 &=&
\mathcal{J}_1^{\bf T}
\left(
 \begin{array}{c}
  \nabla _\mu \frac{1}{T} + \frac{1}{T} D u_\mu \\
  -\nabla _\mu \frac{\mu _1}{T} \\
  -\nabla _\mu \frac{\mu _2}{T} \\
  \cdots
 \end{array}
\right) ,
\\
 \partial _\alpha I_0^{\perp \perp \mu \nu \alpha}
 &=&
 \label{eq:matricesj2}
 \mathcal{J}_2^{\bf T} \frac{1}{T} \nabla ^{\langle \mu} u^{\nu \rangle} .
\end{eqnarray}
Since $\mathcal{C}_i$,
the transport coefficient matrices for $\varepsilon^{\mu \nu}$
and $\varepsilon_J^\mu$, are symmetric,
$\mathcal{A}_i$'s are also completely symmetric
and Onsager reciprocal relations are satisfied.
Also, $\mathcal{A}$'s are semi-positive definite
because $\mathcal{C}$'s are.
These are the linear transformations of the dissipative currents
and of the thermodynamic forces in the entropy production mentioned
in Appendix~\ref{sec:phenom}.
Note that if the distortion of the distribution other than
Eq.~(\ref{eq:dym}) were employed,
the reciprocal relations would not hold
since the moment equations would no longer
be uniquely constrained from the second law of thermodynamics.
Here we emphasize that real hydrodynamic transport coefficients
reflect microscopic physics of the dense medium and are different from the ones obtained in kinetic theory
and that calculation of transport coefficients is
not the aim of the present paper.

It should be noted here that the derivatives of moments
in Eq.~(\ref{eq:ent_pro_IS0}),
$\partial _\alpha I_0^{\mu \nu \alpha}$
and $\partial _\alpha I_{J0}^{\mu \alpha}$,
do not disappear because $f_0$ is the distribution
for \textit{local} thermal equilibrium,
not global one.
These derivatives, as we have seen,
are the sources of Navier-Stokes thermodynamic forces.
The dissipative currents disappear in local thermal equilibrium
because the transport coefficients vanish, not the thermodynamic forces.

\subsection{Energy and Particle Frames}

The constitutive equations for multi-component systems
with multiple conserved currents (\ref{eq:Pi})-(\ref{eq:pi})
are frame independent.
On the other hand, practically speaking, it is convenient to
simplify the constitutive equations without losing generality
by choosing frames.
There are two conventional ways of choosing a frame
in dissipative hydrodynamics:
the energy frame and the particle frame.
They are also known as the Landau frame and the Eckart frame, respectively.
In the energy frame, we set the flow $u^\mu = u^\mu_E$ in the direction
of the flow of energy so that no leak from fluid elements exists,
\textit{i.e.}, $W^\mu=0$.
Note that this implies $T^{\mu}_{\ \nu} u^\nu_E = e_0 u^\mu_E$.
The constitutive equations then reduce to
\begin{eqnarray}
\Pi &=& -\zeta \nabla _\mu u^\mu_E - \tau_\Pi D \Pi \nonumber \\
&+& \sum_J \chi_{\Pi \Pi}^{aJ} \Pi D\frac{\mu _J}{T} 
+ \chi_{\Pi \Pi}^b \Pi D \frac{1}{T} 
+ \chi_{\Pi \Pi}^c \Pi \nabla _\mu u^\mu_E \nonumber \\
&-& \zeta_{\Pi \delta e} D\frac{1}{T}
+ \sum _J \zeta_{\Pi \delta n_J} D\frac{\mu_J}{T} \nonumber \\
&+& \sum_{J,K} \chi_{\Pi V_J}^{aK} V^J_\mu \nabla ^\mu \frac{\mu_{K}}{T}
+ \sum_J \chi_{\Pi V_J}^b V^J_\mu \nabla ^\mu \frac{1}{T} \nonumber \\
&+& \sum_J \chi_{\Pi V_J}^c V^J_\mu D u ^\mu_E 
+ \sum_J \chi_{\Pi V_J}^d \nabla ^\mu V^J_\mu \nonumber \\
&+& \chi_{\Pi \pi} \pi _{\mu \nu} \nabla ^{\langle \mu} u^{\nu \rangle}_E ,
\label{eq:Pi_E}
\end{eqnarray}
\begin{eqnarray}
V_J^\mu &=& \kappa_{V_J} \nabla ^\mu \frac{\mu _J}{T} - \tau _{V_J}
\Delta^{\mu \nu} D V^{J}_\nu \nonumber \\
&+& \sum_{K\neq J} \kappa_{V_J V_K} \nabla ^\mu \frac{\mu _K}{T}
- \sum_{K\neq J} \tau _{V_J V_{K}} \Delta^{\mu \nu} D V^{K}_\nu \nonumber \\
&+& \sum_{K,L} \chi_{V_J V_{K}}^{aL} V_{K}^\mu D \frac{\mu _{L}}{T} \nonumber \\
&+& \sum_{K} \chi_{V_J V_{K}}^b V_{K}^\mu D \frac{1}{T}
+ \sum_{K} \chi_{V_J V_{K}}^c V_{K}^\mu \nabla _\nu u^\nu_E \nonumber \\
&+&  \sum_{K} \chi_{V_J V_{K}}^d V_{K}^\nu \nabla _\nu u^\mu_E 
+ \sum_{K} \chi_{V_J V_{K}}^e V_{K}^\nu \nabla ^\mu u_\nu^E \nonumber \\
&+& \kappa _{V_J W} \bigg( \frac{1}{T} D u_E^\mu
+ \nabla ^\mu \frac{1}{T} \bigg) \nonumber \\
&+& \sum_{K} \chi_{V_J \pi}^{aK} \pi^{\mu \nu} \nabla_\nu \frac{\mu _{K}}{T}
+ \chi_{V_J \pi}^b \pi^{\mu \nu} \nabla_\nu \frac{1}{T} \nonumber \\
&+& \chi_{V_J \pi}^c \pi ^{\mu \nu} D u_\nu^E \nonumber
+ \chi_{V_J \pi}^d \Delta^{\mu \nu} \nabla ^\rho \pi _{\nu \rho} \\
&+& \sum_{K} \chi_{V_J \Pi}^{aK} \Pi \nabla ^\mu \frac{\mu _{K}}{T}
+ \chi_{V_J \Pi}^b \Pi \nabla ^\mu \frac{1}{T} \nonumber \\
&+& \chi_{V_J \Pi}^c \Pi D u^\mu_E + \chi_{V_J \Pi}^d \nabla ^\mu \Pi ,
\label{eq:V_E}
\end{eqnarray}
\begin{eqnarray}
\pi^{\mu \nu} &=& 2 \eta \nabla ^{\langle \mu} u^{\nu \rangle}_E 
- \tau_\pi D \pi^{\langle \mu \nu \rangle} \nonumber \\
&+& \sum_J \chi_{\pi \pi}^{aJ} \pi^{\mu \nu} D \frac{\mu_J}{T} 
+ \chi_{\pi \pi}^b \pi^{\mu \nu} D \frac{1}{T} \nonumber \\
&+& \chi_{\pi \pi}^c \pi ^{\mu \nu} \nabla _\rho u^\rho_E 
+ \chi_{\pi \pi}^d \pi ^{\rho \langle \mu} 
\nabla _\rho u^{\nu \rangle}_E \nonumber \\
&+& \sum_{J,K} \chi_{\pi V_J}^{aJ} V_J^{\langle \mu}
\nabla ^{\nu \rangle} \frac{\mu_{K}}{T} 
+ \sum_J \chi_{\pi V_J}^b V_J^{\langle \mu} 
\nabla ^{\nu \rangle} \frac{1}{T} \nonumber \\
&+& \sum_J \chi_{\pi V_J}^c V_J^{\langle \mu} D u^{\nu \rangle}_E 
+ \sum_J \chi_{\pi V_J}^d \nabla ^{\langle \mu} V_J ^{\nu \rangle} \nonumber \\
&+& \chi _{\pi \Pi} \Pi \nabla ^{\langle \mu} u^{\nu \rangle}_E .
\label{eq:pi_E}
\end{eqnarray}
Note here that the term 
$\kappa _{V_J W} ( \frac{1}{T} D u_E^\mu + \nabla ^\mu \frac{1}{T} )$
in Eq.~(\ref{eq:V_E}) does not vanish
even though it contains the first order thermodynamic force for $W^\mu$,
because the constitutive equations for $W^\mu$
include the term proportional to $\kappa _{W V_J} \nabla^\mu \frac{\mu_J}{T}$
in turn. These phenomena are known as 
Soret effect and Dufour effect respectively
in non-equilibrium statistical mechanics \cite{Prigozine}. 

The particle frame in single conserved current systems
is defined as the frame where no leak of the charge is observed.
Naively this is not well defined
when more than one conserved current is present
because in the frame where $V_J^\mu = 0$,
the other currents would not vanish, \textit{i.e.}, $V_K^\mu \neq 0$
for $J \neq K$.
In this respect we should consider the \textit{average} particle
frame where the sum of the charge dissipation vanishes
in the case of multi-conserved current systems.
We define here the heat current $q^\mu$ in a system 
with $N$ conserved currents as
\begin{equation}
\label{eq:heatcurrent_m}
q^\mu = W^\mu - \sum_J \frac{e_0+P_0}{n_{J0}} V_J^\mu ,
\end{equation}
which corresponds to the energy conduction 
through pure heat conduction
because the contributions from particle diffusions are subtracted.
In the average particle frame one expects $q^\mu = W^\mu$, or equivalently,
$\sum_J \frac{V_J^\mu}{n_J} = 0$.
The flow in this frame can be written as 
\begin{equation}
 u^\mu_N = \frac{1}{N} \sum_J \frac{N_J^\mu}{n_{J0}},
\end{equation}
which is the average of the flows in one-current particle frames.
The resulting constitutive equations have the same tensor structure
as shown in Eqs.~(\ref{eq:Pi})-(\ref{eq:pi})
because each charge current $V_J$ does not vanish.

\section{Discussion}
\label{sec3}

Several comments are in order here.

Firstly, we have derived the constitutive equations
from the law of increasing entropy only.
Actually this should be the case for dissipative hydrodynamic formalism
because it is the only thermodynamic relation
which implies irreversible processes.
This is in good contrast to the fact
that ideal hydrodynamic equations of motion,
which describe time reversible processes,
are the energy-momentum conservation and the charge number conservations.

Secondly, if one decomposed $\delta T^{\mu \nu}$ and $\delta N^\mu _J$
into each component $i$, one would have obtained the equations 
which might be much similar to the single component constitutive equations.
The problems are, however, that
(a) some transport coefficients such as bulk viscosity
cannot be trivially separated into components,
(b) in addition to the constitutive equations
one needs to solve the energy-momentum conservation
and the charge number conservations
but they do not hold for each particle species
and (c) the distortion of the distribution cannot be determined
in such methods without introducing additional microscopic physics,
which often causes lack of generality.
Such constitutive equations would not be equivalent to
the constitutive equations
we obtained.
Moreover, such equations can be
applied only for the system with (quasi-)particle
picture. In general, it is not the case,
in particular, in the vicinity of phase transition
or in highly dense system.

In the following
we would like to investigate other approaches
and discuss the correspondences between these approaches and our formalism.

\subsection{Ambiguities of Second Order Equations
in Phenomenological Approaches}
\label{subsec:3-A}

We calculate the entropy production up to the second order
and investigate the possibility of deriving $10+4N$ second order equations
from the law of increasing entropy by extending the approach mentioned
in Appendix~\ref{sec:phenom}.
We find that second-order equations cannot be uniquely determined 
in this way.
As mentioned before,
the derivation of the second order constitutive equations
requires the information of $\delta f^i$.
If we alternatively expand Eq.~(\ref{eq:entpro_phenom})
up to the second order, we obtain
\begin{eqnarray}
\label{eq:entpro_phenom2}
\partial _\mu s^\mu 
&=& \sum _i \int \frac{g_i d^3 p}{(2\pi )^3 E_i}p_i^\mu
\bigg[ \delta f^i \partial_\mu y_0^i
+ \delta f^i \partial_\mu \delta y^i  \nonumber \\
&+& \frac{1}{2} f_0^i(1\pm f_0^i)(1\pm 2f_0^i)
\delta y^{i2} \partial_\mu y_0^i \bigg] ,
\end{eqnarray}
which is equivalent to Eq. (\ref{eq:entropy}) at this order.
In this case the last term is problematic in obtaining second order equations;
it involves the terms with two dissipative currents coupled 
with one thermodynamic force,
\textit{e.g.}, $\Pi W^\mu \nabla_\mu \frac{1}{T}$ 
when $\delta y^i$ is estimated in the moment expansion.
These terms cannot be naively associated 
with one of the dissipative currents 
to forcefully obtain $10+4N$ equations
because generally the dissipative currents of different tensor structure
can be found in the second order terms,
\textit{e.g.}, $\nabla ^\mu \Pi$ terms
in the equation for $W^\mu$ and $V^\mu$.  

We further consider whether a more phenomenological 
approach \cite{Israel:1979wp, Muronga}, in which one expands 
the entropy current with respect to dissipative currents 
and uses the second law of thermodynamics, yields
full second order constitutive equations. 
If one assumes that the second order distortion 
of entropy current (\ref{eq:ds2}) can be naively
written as the sum of all the possible second order terms
in the dissipative currents, one has
\begin{eqnarray}
\label{eq:ds2_phenom}
\delta s^\mu_{(2)} 
&=& (\alpha_0^{\Pi \Pi} \Pi ^2 + \alpha_0^{\delta e \delta e} \delta e ^2 
+ \sum_{J,K} \alpha_0^{\delta n_J \delta n_{K}}
\delta n_J  \delta n_{K} \nonumber \\
&+& \alpha_0^{\Pi \delta e} \Pi \delta e + \sum_J \alpha_0^{\Pi \delta n_J}
\Pi \delta n_J + \sum_J \alpha_0^{\delta e \delta n_J}
\delta e \delta n_J \nonumber \\
&+& \alpha_0^{W W} W^\nu W_\nu + \sum_J \alpha_0^{W V_J}
W^\nu V^J_\nu \nonumber \\
&+& \sum_{J,K} \alpha_0^{V_J V_{K}} V_J^\nu V^{K}_\nu 
+ \alpha_0^{\pi \pi} \pi^{\nu \rho} \pi_{\nu \rho} ) u^\mu \nonumber \\
&+& \alpha_1^{\Pi W} \Pi W^\mu + \sum_J \alpha_1^{\Pi V_J} \Pi V_J^\mu
+ \alpha_1^{\delta e W} \delta e W^\mu \nonumber \\
&+& \sum_J \alpha_1^{\delta e V_J} \delta e V_J^\mu 
+ \sum_J \alpha_1^{\delta n_J W} \delta n_J W^\mu \nonumber \\
&+& \sum_{J,K} \alpha_1^{\delta n_J V_{K}} \delta n_J V_{K}^\mu
+ \alpha_1^{W \pi} W_\nu \pi^{\mu \nu} \nonumber \\
&+& \sum_J \alpha_1^{V_J \pi} V^J_\nu \pi^{\mu \nu} ,
\end{eqnarray}
where $\alpha$'s are coefficients.
Again the derivative of the entropy current involves the terms
with two dissipative currents coupled with one thermodynamic gradient,
so we cannot determine the second order constitutive equations in this way.
Such terms are naively dropped in Ref.~\cite{Israel:1979wp}. 

It is note-worthy that if one considers kinetic theory with 
our extended Grad's moment method to estimate Eq.~(\ref{eq:ds2}), 
the resulting entropy current will have the same tensor structure 
as Eq.~(\ref{eq:ds2_phenom}) does.
In this case the coefficients $\alpha$'s are fixed in kinetic theory.
When one writes down the entropy current first 
then takes its derivative this way, 
however, the terms proportional to
$\Pi \pi^{\mu \nu}$ do not seem to appear in $s^\mu$ and 
consequently in the constitutive equations
since the only possible way to construct Lorentz vector
from bulk-shear term at the second order is $\Pi \pi^{\mu \nu} u_\nu = 0$.
This does not contradict our results 
because there remains the ambiguity 
that we can add arbitrary amount of
$\Pi \pi^{\mu \nu} \partial _\mu u_\nu - \pi^{\mu \nu}
\Pi \partial _\mu u_\nu (= 0)$
to the entropy production, and associate one term with
the constitutive equation for the bulk pressure
and the other with the ones for the shear stress tensor.
This corresponds to formally keeping 
$\alpha_1^{\Pi \pi} \Pi \pi^{\mu \nu} u_\nu (=0)$
in the entropy current.
In other words, constitutive equations from kinetic theory 
in general can have the bulk-shear terms, although most 
of the conventional formalisms seem to be unaware of it, 
possibly because they cannot determine the amount of such terms.

Thus the constitutive equations cannot be 
uniquely determined by the phenomenological approach, 
nor by simply taking derivative of the expansion of 
the entropy current (\ref{eq:ds2}) in kinetic theory.
The ambiguity of associating terms to constitutive equations,
of course, is removed in our extended Israel-Stewart formalism.

\subsection{Single-Component Systems without Chemical Interaction}
\label{subsec:3-B}

In the conventional second order theories,
including Israel-Stewart theory, only the moment equations
$\partial_\alpha I^{\mu \nu \alpha} = Y^{\mu \nu}$ (\ref{eq:third_multi})
are considered.
Naively speaking the number of equations is 10 in this case,
but if the single component systems with no particle production
nor annihilation are assumed the particle number current $N^\mu$
becomes the conserved current, and the number of equations is reduced to 9.
This is because the trace of the moment equations
$g_{\mu \nu} \partial _\alpha I^{\mu \nu \alpha}
= g_{\mu \nu} Y^{\mu \nu}$ coincides with
the number conservation $\partial_\mu N^\mu = 0$, \textit{i.e.},
$\partial _\alpha I^{\ \mu \alpha}_\mu = m^2 \partial_\mu N^\mu = 0$
in kinetic theory. The two scalar equations
$u_\mu u_\nu \partial_\alpha I^{\mu \nu \alpha} = u_\mu u_\nu Y^{\mu \nu}$
and $\Delta_{\mu \nu} \partial_\alpha I^{\mu \nu \alpha}
= \Delta_{\mu \nu} Y^{\mu \nu}$ then become identical.
On the other hand, the number of dissipative currents
is 14 but can also be reduced to 9 by using the stability conditions
$\delta e = \delta n = 0$ and choosing the frame
to drop 3 dissipative currents from either $W^\mu$ or $V^\mu$.
This means $\Pi$, $q^\mu$ and $\pi^{\mu \nu}$ are considered.
Here the heat current $q^\mu$ in single component systems is defined as 
\begin{equation}
\label{eq:heatcurrent}
q^\mu = W^\mu - \frac{e_0+P_0}{n_{0}} V^\mu ,
\end{equation}
which reduces to either $W^\mu$ or $V^\mu$
depending on whether one chooses the particle frame or the energy frame.
However, the number of equations and that of dissipative currents
no longer match in multi-component systems, because as mentioned earlier
\begin{equation}
\label{eq:trace2}
g_{\mu \nu} \partial _\alpha I^{\mu \nu \alpha} 
= \sum _ i m_i ^2 \int \frac{g_i d^3 p}{(2\pi )^3 E_i} p_i^\alpha
\partial _\alpha f^i \neq 0 ,
\end{equation}
is not a conserved current even if no particle production nor annihilation is assumed. In a system with particle creations and annihilations
which we consider in this paper,
what one really has is the charge number conservations
$\partial _\mu N^{\mu}_J = 0$.
Thus we have 10 moment equations
in multi-component systems.
This means that naive generalization of the conventional second order theory to multi-component systems does not work because only 9 dissipative currents are considered there. Also if the system has more than one conserved current, the number of vector dissipative currents exceeds that of equations even if the frame is fixed.

The apparent inconsistency between the number of equations and that of unknowns arise from the three facts. 
Firstly, the stability conditions $\delta e = \delta n_J = 0$ are employed prematurely and $1+N$ unknowns are omitted. The stability conditions are employed to ensure that the system is in maximum entropy state, $i.e.$, thermodynamically stable, and thus are different physics from the count of the number of unknowns in kinetic theory. In this sense the stability conditions have to be considered after the constitutive equations are derived. 
Secondly, the moment equations $\partial_\alpha I_J^{\mu \alpha} = Y_J^{\mu}$ (\ref{eq:third_multi_b}) are not considered for the systems with conserved currents. This means that $4N$ equations are missing in the formalism. 
Thirdly, in the conventional approach, one considers the heat current $q^\mu$ instead of the energy current $W^\mu$ and the charge currents $V_J^\mu$, possibly due to the limitation of the number of equations. This means $3$ out of $3+3N$ vector dissipative currents are taken into account. Thus it is impossible to determine all the vector dissipative currents $W^\mu$ and $V_J^\mu$ simultaneously in an arbitrary frame within the conventional Israel-Stewart approach. Related discussion can be found in Appendix~\ref{mcc}.

\subsection{Correspondences with Other Formalisms}
\label{subsec:3-C}

We would like to discuss correspondences between our formalism and other frameworks. It should be noted that naive comparisons cannot be made because all the other formalisms are derived in systems with single component and/or single conserved current and are sometimes frame dependent. Therefore we take specific conditions such as the energy frame or the particle frame in the single conserved current limit to make the correspondences clear.

\subsubsection{Constitutive Equations from AdS/CFT}

We compare our results with the conformal equations for $\pi^{\mu \nu}$ based on Anti de-Sitter Space/Conformal Field Theory (AdS/CFT) correspondence
\cite{Baier:2007ix}.
Our constitutive equations for the shear stress tensor (\ref{eq:pi}) can be expressed as, in the conformal limit $\Pi=0$,
\begin{eqnarray}
\pi^{\mu \nu} &=& 2 \eta \nabla ^{\langle \mu} u^{\nu \rangle} - \tau_\pi D \pi^{\langle \mu \nu \rangle} + \chi_{\pi \pi}^d \pi^{\rho \langle \mu} \omega ^{\nu \rangle} _{\ \rho} \nonumber \\
&+& \chi_{\pi \pi}^d \pi ^{\rho \langle \mu} \sigma ^{\nu \rangle}_{\ \rho} + \bigg( \frac{1}{3} \chi_{\pi \pi}^d + \chi_{\pi \pi}^c \bigg) \pi ^{\mu \nu} \nabla _\rho u^\rho \nonumber \\
&+& \chi_{\pi \pi}^b \pi^{\mu \nu} D \frac{1}{T} ,
\label{eq:pi_cl}
\end{eqnarray}
when estimated in the energy frame and in the zero net charge density limits. The identity (\ref{eq:tensor_vortisity_indentity}) is used here. The first term is the Navier-Stokes term, the following two terms are the conventional second order terms, and the terms in the second line are the new terms that appear when the consistent expansion is performed as mentioned in the previous section.
It is note-worthy that the term proportional to $\Pi \nabla ^{\langle \mu} u^{\nu \rangle}$ is omitted in the conformal limit but is equivalent to the $\pi^{\mu \nu} \nabla_\rho u^\rho$ term in second order theory, because both reduces to $-\zeta \mathrm{(or \ 2 \eta)} \nabla ^{\langle \mu} u^{\nu \rangle} \nabla_\rho u^\rho$ when the first order expressions are utilized. 
The former vanishes while the latter does not, because the transport coefficient $\zeta$ vanishes in the conformal limit, not the thermodynamic force itself.

On the other hand, the constitutive equations from Ref.~\cite{Baier:2007ix} in flat space are
\begin{eqnarray}
\label{eq:shear_adscft}
\pi^{\mu \nu} &=& 2 \eta \nabla ^{\langle \mu} u^{\nu \rangle} - \tau_\pi D \pi^{\langle \mu \nu \rangle} - \frac{d}{d-1} \tau_\pi \pi ^{\mu \nu} \nabla_\rho u^\rho \nonumber \\
&+& \frac{\lambda _1}{\eta ^2} \pi ^{\rho \langle \mu} \pi ^{\nu \rangle} _{\ \rho} - \frac{\lambda _2}{\eta} \pi ^{\rho \langle \mu} \omega ^{\nu \rangle} _{\ \rho} + \lambda _3 \omega ^{\rho \langle \mu} \omega ^{\nu \rangle} _{\ \rho} ,
\end{eqnarray}
where notations are adjusted to our formalism. This holds for single component systems, but direct comparison with our formalism can be made because the constitutive equations of the shear stress tensor is free from the non-trivialities of multi-component systems. Note that the terms proportional to $\pi^{\mu \nu} D \frac{1}{T}$ in Eq.~(\ref{eq:pi_cl}) can be absorbed in the $\pi ^{\mu \nu} \nabla _\rho u^\rho$ term, because of the ideal hydrodynamic relation
\begin{eqnarray}
TD\frac{1}{T} &=& \bigg( \frac{\partial P_0}{\partial e_0} \bigg) _{n_0} \nabla _\mu u^\mu .
\label{eq:ideal_relations_beta}
\end{eqnarray}
Then obvious correspondences can be found for all the terms, except for the $\omega ^{\rho \langle \mu} \omega ^{\nu \rangle} _{\ \rho}$ term; the terms with no dissipative currents generally do not appear in our formalism, because the distribution is expanded in terms of the dissipative currents with Grad's moment method. 
The AdS/CFT approach \cite{Baier:2007ix} yields that term because all the possible terms which are consistent with their approach are added manually in the derivation. If one added the vorticity-vorticity term phenomenologically in our formalism, that would be inconsistent in the context of the derivation in kinetic theory.

\subsubsection{Constitutive Equations from Renormalization Group Method}

Next we investigate the correspondences between the constitutive
equations obtained in this paper
and the ones from renormalization group approach \cite{Tsumura:2007ji}.
Here we consider a single conserved current system.
In the energy frame the constitutive equations from that paper are 
\begin{eqnarray}
\Pi &=& -\zeta \nabla _\mu u^\mu - \tau _\Pi D \Pi \nonumber \\
&+& \tau_\Pi \bigg[ - \frac{1}{2} \frac{T \zeta}{\tau_\Pi} \partial_\mu \bigg( \frac{\tau_\Pi u^\mu}{T \zeta} \bigg) \nonumber \\
&+& \frac{1}{2} \bigg(- D\frac{\mu}{T} + T \delta^{(0)}_\Pi D\frac{1}{T} + \delta^{(1)}_\Pi \nabla_\mu u^\mu \bigg) \bigg] \Pi \nonumber \\
&+& l_{\Pi V} \bigg[ - \nabla_\mu \frac{\mu}{T} + T \delta_{\Pi V} \bigg( \nabla_\mu \frac{1}{T} + \frac{1}{T} D u_\mu \bigg) \bigg] V^\mu \nonumber \\
&-& l_{\Pi V} \nabla _\mu V^\mu + l_{\Pi \pi} \nabla_{\langle \mu} u_{\nu \rangle} \pi^{\mu \nu} ,
\label{eq:Pi_TKO}
\end{eqnarray}
\begin{eqnarray}
V^\mu &=& \kappa \bigg( \frac{n_0 T}{e_0 +P_0} \bigg)^2 \nabla^\mu \frac{\mu}{T} - \tau_V \Delta^{\mu}_{\ \nu} D V^\nu \nonumber \\
&+& \tau_V \bigg\{ - \frac{1}{2} \frac{\kappa}{\tau_V} \bigg( \frac{n_0 T}{e_0 +P_0} \bigg)^2 \partial_\nu \bigg[ \frac{\tau_V u^\nu}{\kappa} \bigg( \frac{e_0 +P_0}{n_0 T} \bigg)^2 \bigg] \nonumber \\
&+& \frac{1}{2} \bigg( - D\frac{\mu}{T} + T \delta^{(0)}_{V} D \frac{1}{T} + \frac{5}{3} \delta^{(1)}_V \nabla_\nu u^\nu \bigg) \bigg\} V^\mu \nonumber \\
&+& \tau_V \delta^{(1)}_V 2 \nabla^{\langle \mu} u^{\nu \rangle} V_\nu \nonumber \\
&+& l_{V \pi} \bigg[ - \nabla_\nu \frac{\mu}{T} + T \delta_{V \pi} \Big( \nabla_\nu \frac{1}{T} + \frac{1}{T} Du_\nu \Big) \bigg] \pi ^{\mu \nu} \nonumber \\
&-& l_{V \pi} \nabla_\nu \pi ^{\langle \mu \nu \rangle} \nonumber \\
&+& l_{V \Pi} \bigg[ -\nabla^\mu \frac{\mu}{T} + T \delta_{V \Pi} \bigg( \nabla^\mu \frac{1}{T} + \frac{1}{T} D u^\mu \bigg) \bigg] \Pi \nonumber \\
&-& l_{V \Pi} \nabla^\mu \Pi ,
\label{eq:V_TKO}
\end{eqnarray}
\begin{eqnarray}
\pi^{\mu \nu} &=& 2 \eta \nabla^{\langle \mu} u^{\nu \rangle} - \tau_\pi D \pi^{\langle \mu \nu \rangle} \nonumber \\
&+& \tau_\pi \bigg[ - \frac{1}{2} \frac{T \eta}{\tau_\pi} \partial_\rho \bigg( \frac{\tau_\pi  u^\rho}{T  \eta} \bigg) \nonumber \\
&+& \frac{1}{2} \bigg( - D\frac{\mu}{T} + T \delta^{(0)}_{\pi} D \frac{1}{T} + \frac{7}{3} \delta^{(1)}_\pi \nabla_\rho u^\rho \bigg) \bigg] \pi^{\mu \nu} \nonumber \\
&+& \tau_\pi \delta^{(1)}_\pi 4 \pi^{\rho \langle \mu} \sigma^{\nu \rangle}_{\ \rho} \nonumber \\
&+& l_{\pi V} \bigg[ - \nabla^{\langle \mu} \frac{\mu}{T} + T \delta_{\pi V}  \bigg( \nabla^{\langle \mu} \frac{1}{T} + \frac{1}{T} D u^{\langle \mu} \bigg) \bigg] V^{\nu \rangle} \nonumber \\
&-& l_{\pi V} \nabla^{\langle \mu} V^{\nu \rangle} + l_{\pi\Pi} \nabla^{\langle \mu} u^{\nu \rangle} \Pi .
\label{eq:pi_TKO}
\end{eqnarray}
Our formalism includes the vorticity terms $\omega^{\mu \nu} V_\nu$ and $\pi^{\rho \langle \mu} \omega^{\nu \rangle}_{\ \rho}$ whereas Eqs. (\ref{eq:V_TKO}) and (\ref{eq:pi_TKO}) do not. On the other hand the terms which involve derivatives of transport coefficients do not exist in our formalism, because such terms cannot be expressed with dissipative currents, unless we assume kinetic theory for the relaxation times over viscosities, \textit{e.g.}, $\frac{\tau_\Pi}{\zeta}$, and express them in terms of $J_{mn}$'s. 
Here we note that, as is the case for ordinary hydrodynamics, these coefficients depend on space-time coordinates through their temperature and chemical potential dependences in our formalism.
Aside from the differences accounted for non-trivialities of multi-component systems, the equations have almost the same tensor structure as that in our formalism with different coefficients. The similarity in tensor structure is worth-mentioning, considering that their formalism uses the technique based on renormalization group theory and thus is different from our formalism.
It should be mentioned that this formalism has frame dependence, and the forms of the constitutive equations differ from the above ones when the particle frame is employed. 

\subsubsection{Constitutive Equations from Methods of 14 Moments}

The second order single-component constitutive equations in the original Israel-Stewart formalism \cite{Israel:1979wp} are
\begin{eqnarray}
\Pi &=& - \zeta \partial _\mu u^\mu_E - \zeta \beta_0 D \Pi \nonumber \\
&+& \zeta a^\prime_0 q^\mu  D u^\mu_E + \zeta \alpha_0 \partial_\mu q^\mu ,
\label{eq:Pi_IS}
\end{eqnarray}
\begin{eqnarray}
q^\mu &=& \kappa T \frac{n_0 T}{e_0 +P_0} \nabla^\mu \frac{\mu}{T} - \kappa T \beta_1 \Delta^{\mu \nu} D q_\nu \nonumber \\
&+& \kappa T \beta_1 \omega^{\mu \nu} q_\nu \nonumber \\
&+& \kappa T a_1 \pi^{\mu \nu} D u_\nu^E + \kappa T \alpha_1 \Delta^{\mu \nu} \partial ^\rho \pi_{\nu \rho } \nonumber \\
&+& \kappa T a_0 \Pi D u^\mu_E + \kappa T \alpha_0 \nabla^\mu \Pi ,
\label{eq:q_IS}
\end{eqnarray}
\begin{eqnarray}
\pi^{\mu \nu} &=& - 2 \eta \nabla^{\langle \mu} u^{\nu \rangle}_E - 2 \eta \beta_2 D \pi^{\langle \mu \nu \rangle} \nonumber \\
&+& 4 \eta \beta _2 \pi_\rho^{\ \langle \mu} \omega^{\nu \rangle \rho} \nonumber \\
&+& 2 \eta a_1^\prime q^{\langle \mu} D u^{\nu \rangle} + 2 \eta a_1 \partial^{\langle \mu} q^{\nu \rangle}.
\label{eq:pi_IS}
\end{eqnarray}
Note that the metric used in the above paper is opposite to ours, which is the source of the negative sign before $\eta$ in the first term in Eq.~(\ref{eq:pi_IS}). Here single conserved current systems with binary collisions are considered. Compared with our formalism, they discards the second order terms with respect to the first order thermodynamic forces, \textit{i.e.}, $D\frac{1}{T}$, $D\frac{\mu}{T}$, $\nabla_\mu u^\mu$, $\nabla_\mu \frac{1}{T}$, $\nabla_\mu \frac{\mu}{T}$, and $\nabla_{\langle \mu} u_{\nu \rangle}$, stating they would be small. It is worth-mentioning, however, that the terms proportional to acceleration $Du_\mu$ are kept in the equations.

Several post Israel-Stewart second-order terms are found for single component systems in Ref.~\cite{Muronga} by phenomenologically expressing the entropy currents in terms of the dissipative currents up to the second order. The constitutive equations are, in the particle frame,
\begin{eqnarray}
\Pi &=& - \zeta \partial _\mu u^\mu - \tau_\Pi D \Pi \nonumber \\
&-& \frac{1}{2} \zeta T \partial_\mu \bigg( \frac{\tau_0 u^\mu}{\zeta T} \bigg) \Pi + \tau_0 \nabla_\mu q^\mu ,
\label{eq:Pi_M}
\end{eqnarray}
\begin{eqnarray}
q^\mu &=& \kappa ( \nabla^\mu T - T D u^\mu ) - \tau_q \Delta^{\mu \nu} D q_\nu \nonumber \\
&+& \frac{1}{2} \kappa T^2 \partial_\nu \bigg( \frac{\tau u^\nu}{\kappa T^2} \bigg)q^\mu \nonumber \\
&-& \tau_1 \nabla_\nu \pi^{\mu \nu} - \tau_0 \nabla^\mu \Pi ,
\label{eq:q_M}
\end{eqnarray}
\begin{eqnarray}
\pi^{\mu \nu} &=& 2 \eta \nabla^{\langle \mu} u^{\nu \rangle} - \tau_\pi D \pi^{\langle \mu \nu \rangle} \nonumber \\
&-& \eta T \partial_\lambda \bigg( \frac{\tau_2 u^\lambda}{2 \eta T} \bigg) \pi^{\mu \nu} + \tau_2 \nabla^{\langle \mu} q^{\nu \rangle}.
\label{eq:pi_M}
\end{eqnarray}
The new terms here are the ones proportional to $\Pi \nabla_\mu u^\mu$, $q^\mu \nabla_\nu u^\nu$ and $\pi^{\mu \nu} \nabla_\rho u^\rho$. Terms which involve acceleration $Du^\mu$ do not appear in the second order terms. It is also mentioned in the paper that when the kinetic approach is employed, vorticity terms $\tau_q \omega^{\mu \nu} q_\nu$ and $\tau_\pi \pi_\lambda^{\ \langle \mu} \omega^{\nu \rangle \lambda}$ appear in Eq.~(\ref{eq:q_M}) and in Eq.~(\ref{eq:pi_M}), respectively. Note that due to the ambiguities of the phenomenological approaches mentioned in Sec.~\ref{subsec:3-A}, bulk pressure $\Pi$ does not appear in the constitutive equations of shear stress tensor $\pi^{\mu \nu}$, and vice versa. 

More second order terms are reported in Ref. \cite{Betz:2009zz} in the framework of Israel-Stewart theory with Grad's 14-moment method. 
Actually when our formalism is reduced to single conserved current systems and the non-trivialities of the multi-component systems are omitted,
the third moment equations from the two formalisms should become equivalent except for transport coefficients 
because both of them follow Israel-Stewart theory consistently.
The apparent difference from our formalism comes from expansions. They expand $I^{\mu \nu \alpha}$ first and then take the derivative
\begin{eqnarray}
\partial _\alpha I^{\mu \nu \alpha} &=& \partial _\alpha I_0^{\mu \nu \alpha} - \partial _\alpha ({J^{\mu \nu \alpha} \varepsilon}) \nonumber \\
&-& \partial _\alpha (J^{\mu \nu \alpha \beta} \varepsilon _{\beta}) - \partial _\alpha (J^{\mu \nu \alpha \beta \gamma} \tilde{\varepsilon}_{\beta \gamma}),
\label{eq:BHR_1}
\end{eqnarray}
whereas we take the derivative first, then expand it as shown in Eq.~(\ref{eq:lhs}). Here $\tilde{\varepsilon}_{\beta \gamma} = \varepsilon_{\beta \gamma} - \frac{\mathrm{Tr}(\varepsilon_{\beta \gamma})}{4} g_{\beta \gamma}$ is a traceless tensor. The coefficient $\varepsilon$ can be identified with $\frac{\mathrm{Tr}(\varepsilon_{\beta \gamma})}{4}$ in single component systems.

The two expansions should yield the same results up to the second order in small quantities, because both of them expand the derivative of the distribution as
\begin{eqnarray}
\partial _\alpha f &=& - f_0 (1\pm f_0) \partial _\alpha y_0 - f_0 (1\pm f_0) \partial _\alpha \delta y \nonumber \\
&+& f_0 (1\pm f_0)(1\pm 2f_0) \delta y \partial_\alpha y_0 .
\label{eq:cons_exp}
\end{eqnarray}
The first term of the right hand side corresponds to the Navier-Stokes limit. The second term is the source of the Israel-Stewart second order terms which includes the derivatives of the dissipative currents. Note that all the terms derived from this second term do not appear in the original paper by Israel and Stewart \cite{Israel:1979wp} as mentioned earlier. The third term corresponds to ``new" terms that are not shown in Ref. \cite{Israel:1979wp}. These terms cannot be neglected because they are also second order terms.

We again consider the energy frame to compare our multi-component results with single component ones, \textit{i.e.}, $q^\mu = - \frac{n_0}{e_0 +P_0} V^\mu$. According to Ref.~\cite{Betz:2009zz}, the constitutive equations are
\begin{eqnarray}
\Pi &=& -\zeta \nabla _\mu u^\mu - \tau_\Pi D \Pi \nonumber \\
&-& \zeta \hat{\delta}_0 \Pi \nabla _\mu u^\mu + \tau_{\Pi q} q_\mu D u^\mu - l_{\Pi q} \partial_\mu q^\mu \nonumber \\
&+& \lambda_{\Pi q} q_\mu \nabla^\mu \frac{\mu}{T} + \lambda_{\Pi \pi} \pi_{\mu \nu} \nabla^{\langle \mu} u^{\nu \rangle} ,
\label{eq:Pi_BHR}
\end{eqnarray}
\begin{eqnarray}
q^\mu &=& \kappa T \frac{n_0 T}{e_0 + P_0} \nabla^\mu \frac{\mu}{T} - \tau_q \Delta^{\mu \nu} D q_\nu \nonumber \\
&-& \kappa T \hat{\delta}_1 q^\mu \nabla _\nu u^\nu - \lambda_{qq} \nabla^{\langle \mu} u^{\nu \rangle} q_\nu + \tau_q \omega^{\mu \nu} q_\nu \nonumber \\
&+& \lambda_{q \pi} \pi^{\mu \nu} \nabla_\nu \frac{\mu}{T} - \tau_{q \Pi} \pi^{\mu \nu} D u_\nu - l_{q \pi} \Delta^{\mu \nu} \partial^\lambda \pi_{\nu \lambda} \nonumber \\
&+& \lambda_{q \Pi} \Pi \nabla^\mu \frac{\mu}{T} + \tau_{q \Pi} \Pi D u^\mu + l_{q \Pi} \nabla^\mu \Pi  ,
\label{eq:q_BHR}
\end{eqnarray}
\begin{eqnarray}
\pi^{\mu \nu} &=& 2\eta \nabla^{\langle \mu} u^{\nu \rangle} - \tau_\pi D \pi^{\langle \mu \nu \rangle} \nonumber \\
&-& 2 \eta \hat{\delta}_2 \pi^{\mu \nu} \nabla _\lambda u^\lambda -2 \tau_\pi \pi_\lambda^{\ \langle \mu} \sigma^{\nu \rangle \lambda} + 2 \tau_\pi \pi_\lambda^{\ \langle \mu} \omega^{\nu \rangle \lambda} \nonumber \\
&-& 2 \lambda_{\pi q} q^{\langle \mu} \nabla^{\nu \rangle} \frac{\mu}{T} + 2 \tau_{\pi q} q^{\langle \mu} D u^{\nu \rangle} + 2 l_{\pi q} \partial^{\langle \mu} q^{\nu \rangle} \nonumber \\
&+& 2 \lambda_{\pi \Pi} \Pi \nabla^{\langle \mu} u^{\nu \rangle} ,
\label{eq:pi_BHR}
\end{eqnarray}
which should be identified with our constitutive equations by taking into account the ideal hydrodynamic relations (\ref{eq:ideal_relations_beta}) and
\begin{eqnarray}
TD\frac{\mu }{T} &=& - \bigg( \frac{\partial P_0}{\partial n_{0}} \bigg) _{e_0} \nabla _\mu u^\mu ,
\label{eq:ideal_relations_alpha} \\
\nabla^\mu \frac{1}{T} + \frac{1}{T} D u^\mu &=& \frac{n_0}{e_0 +P_0} \nabla^\mu \frac{\mu}{T} ,
\label{eq:ideal_relations_vector}
\end{eqnarray}
and the identities (\ref{eq:tensor_vortisity_indentity}) and (\ref{eq:vector_vortisity_indentity2}).

A generalization of Israel-Stewart theory to a relativistic gas mixture is investigated in Ref.~\cite{Prakash:1993bt}. Their formalism is different from ours in several aspects. The systems with no particle creation or annihilation are considered in that paper, \textit{i.e.}, the numbers of each particle species are conserved, while we consider the systems with conservations based on quantum numbers. Their derivation of the equations of motion explicitly depends on the Boltzmann equation, and the dissipative currents are split for each components. On the other hand, our formalism aims derivation of the dissipative hydrodynamic equations and dissipative currents cannot be split.
In the single component limit that formalism reduces to the original Israel-Stewart formalism without acceleration $Du^\mu$ and vorticity $\omega^{\mu \nu}$.

\section{Conclusions}
\label{sec4}

We have derived second order constitutive equations in multi-component systems with multiple conserved currents. Several new second order terms which do not appear in the original Israel-Stewart theory 
are obtained. We found that naive generalization of the conventional Israel-Stewart theory fails due to the mismatching of the number of equations and that of dissipative currents. Several non-trivialities have to be taken into account for consistent derivation of the constitutive equations. Firstly, one must employ the stability condition after the derivation of the constitutive equations. This is important because if the conditions were employed beforehand, the correspondences between the numbers of equations and unknowns would be lost and the system could be described only when sets of assumptions are made, as one can see in the case of single component systems with one conserved current and no chemical interaction. Secondly, in systems with conserved currents, additional moment equations, which are the second moments of the distribution $f^i$ with conserved charges as weight factors, should be introduced. This allows us to describe the system completely. 
Thirdly, Grad's moment method for the determination of the distortion of the distribution, which is necessary for the derivation of the second order constitutive equations, should also be modified to match the number of equations and that of unknowns. We generalized the moment method so that the resulting constitutive equations satisfy Onsager reciprocal relations.
Fourthly, the law of increasing entropy requires the correction tensor to the distribution $\varepsilon_{\mu \nu}$ to have finite trace in multi-component systems.

We explicitly calculated the entropy production in kinetic theory and made clear that the extended Israel-Stewart theory with the moment method indeed satisfies in the first order limit the Onsager reciprocal relations which demand the transport coefficient matrices to be symmetric. It is important to note that all thermodynamic forces of the same tensor order appear in the constitutive equation for a dissipative current. For example, the charge diffusion due to the spatial gradient in temperature is known as Soret effect and energy dissipation due to the spatial gradient in chemical potential as Dufour effect. We further investigated other formalisms, and the phenomenological approaches are found to be unsuitable for the derivation of the second order equations because they include ambiguities in associating second order terms to the constitutive equations.
 Our second order multi-component equations mostly agree with the equations of other formalisms in the single conserved current limit, except for the transport coefficients and for the presence of independent equations of $W^\mu$ and $V_J^\mu$ in our formalism.
 
\acknowledgments
The authors acknowledge fruitful discussions with 
T.~Hatsuda, T.~Kodama, T.~Koide, T.~Kunihiro and S.~Muroya.
The work of T.H. was partly supported by
Grant-in-Aid for Scientific Research
No.~19740130 and by Sumitomo Foundation
No.~080734.  

\appendix

\section{First Order Dissipative Hydrodynamics
for Multi-Component Systems with Multiple Conserved Currents}
\label{sec:phenom}

We briefly review first order dissipative hydrodynamics
for relativistic multi-component systems with multiple conserved currents
with emphasis on the Onsager reciprocal relation.
The entropy current is expanded in terms of
$\delta f^i = f^i - f_0^i$ as
\begin{eqnarray}
\label{eq:ent_exp}
s^\mu &=& s_0^\mu + \sum _i \int \frac{g_i d^3 p}{(2\pi )^3 E_i} p_i^\mu y_0^i \delta f^i + \mathcal{O}(\delta f^2) ,
\end{eqnarray}
up to the first order.
Then the entropy production is
\begin{eqnarray}
\label{eq:entpro_phenom}
\partial _\mu s^\mu &=& \sum _i \int \frac{g_i d^3 p}{(2\pi )^3 E_i}
p_i^\mu \delta f^i \partial_\mu y_0^i
+ \mathcal{O}[\partial (\delta f^2)] \nonumber \\
&\approx & \delta T^{\mu \nu} \partial _\mu \frac{u_\nu}{T}
-  \sum_J \delta N_J^\mu \partial_\mu \frac{\mu_J}{T} \nonumber \\
&=& \delta e D \frac{1}{T} - \Pi \frac{1}{T} \nabla_\mu u^\mu
+ W^\mu \bigg( \nabla _\mu \frac{1}{T}
+ \frac{1}{T}D u_\mu \bigg) \nonumber \\
&+& \pi^{\mu \nu} \frac{1}{T} \nabla _{\langle \mu} u_{\nu \rangle}
- \sum_J \delta n_J D \frac{\mu _J}{T} \nonumber \\
&-& \sum_J V_J^\mu \nabla _\mu \frac{\mu_J}{T}.
\end{eqnarray}
Here we have used the energy-momentum conservation
and the charge number conservations for the first equality.
Then one finds that the dissipative currents should be expressed as 
\begin{eqnarray}
\delta e &=& \zeta_{\delta e \delta e} D\frac{1}{T} - 
\sum_J \zeta_{\delta e \delta n_J} D \frac{\mu_J}{T} \nonumber \\
&+& \zeta_{\delta e \Pi} \frac{1}{T} \nabla_\mu u^\mu , \\
\Pi &=& -\zeta_{\Pi \Pi} \frac{1}{T} \nabla_\mu u^\mu -
\zeta_{\Pi \delta e} D\frac{1}{T} \nonumber \\
&+& \sum_J \zeta_{\Pi \delta n_J} D \frac{\mu_J}{T} , \\
W^\mu &=& -\kappa _{W W} \bigg( \nabla ^\mu \frac{1}{T}
+ \frac{1}{T}D u^\mu \bigg) \nonumber \\
&+& \sum_J \kappa _{W V_J} \nabla ^\mu \frac{\mu_J}{T} , \\
\pi^{\mu \nu} &=& 2 \eta_{\pi \pi} \frac{1}{T} 
\nabla ^{\langle \mu} u^{\nu \rangle}, \\
\delta n_J &=& - \sum_{K} \zeta_{\delta n_J \delta n_K}
D \frac{\mu_K}{T} + \zeta_{\delta n_J \Pi} \frac{1}{T} 
\nabla_\mu u^\mu \nonumber \\
&+& \zeta_{\delta n_J \delta e} D\frac{1}{T} , \\
V_J^\mu &=& \sum_{K} \kappa _{V_J V_K} \nabla ^\mu \frac{\mu_K}{T} \nonumber \\
&-& \kappa _{V_J W} \bigg( \nabla ^\mu \frac{1}{T} + \frac{1}{T}D u^\mu \bigg) , 
\end{eqnarray}
in linear response theory. 
$\zeta = \frac{\zeta_{\Pi \Pi}}{T}$ is bulk viscosity,
$\kappa _W = \kappa_{WW}$ energy conductivity,
$\eta = \frac{\eta_{\pi \pi}}{T}$ shear viscosity
and $\kappa _{V_J} = \kappa_{V_J V_J}$ charge conductivity
of the $J$-th conserved current.
Note that different thermodynamic forces of
the same tensor order are allowed 
in constitutive equations,
such as the gradient of $\frac{\mu_J}{T}$
in the constitutive equation for $W^\mu$.
For such ``cross terms",
the transport coefficients should satisfy
Onsager reciprocal relations \cite{Onsager}, \textit{e.g.},
$\kappa_{WV_J} = \kappa_{V_J W}$ and have to be so chosen
that semi-positive definiteness of the entropy production is preserved. 

One might think each term in Eq.~(\ref{eq:entpro_phenom}) should be expressed
as a quadratic form in dissipative currents
to obey the second law of thermodynamics,
\textit{e.g.}, $W^\mu \propto ( \nabla _\mu \frac{1}{T} + \frac{1}{T}D u_\mu)$
and some of the cross terms seemingly violate it
when Onsager reciprocal relation is considered.
However we can choose the coefficients so that
the law of increasing entropy is actually satisfied.
If we denote the dissipative currents of the same tensor order
as $J_p$ and the corresponding thermodynamic forces
as $X_p$, then the entropy production for this tensor order is
\begin{equation}
\partial _\mu s^\mu = \sum_p J_p X_p.
\end{equation}
The dissipative currents in linear response theory
are expressed as $J_p = \sum_q C_{pq} X_q$
where $C_{pq} = C_{qp}$ is an element of the transport coefficient matrix.
Thus we have
\begin{equation}
\partial _\mu s^\mu = \sum_{p,q} X_p C_{pq} X_q.
\end{equation}
On the other hand, one can diagonalize any symmetric matrix
with a certain orthogonal matrix $P^{-1} = P^{\bf T}$
as $C_{pq} = \sum_{r,s} P^{\bf}_{pr} (C^\prime_r \delta_{rs}) P_{sq}$.
Here one requires the diagonal elements $C^{\prime}_{r}$ must be positive.
Then 
\begin{eqnarray}
\partial _\mu s^\mu &=& \sum_{p,q,r,s} X_p P^{\bf}_{pr} (C^\prime_r \delta_{rs}) P_{sq} X_q \nonumber \\
&=& \sum_{r,s} X_r^\prime (C^\prime_r \delta_{rs}) X_s^\prime \nonumber \\
&=& \sum_r C_r^\prime X_r^{\prime 2} \geq 0 ,
\end{eqnarray}
is obtained where $X_p^\prime = \sum_q P_{pq} X_q$.
It means that if the appropriate linear combinations of
the thermodynamic forces are taken,
the entropy production can be expressed
in quadratic forms, which obviously satisfies
its semi-positive definiteness.

It is important to mention that the constitutive equations
for $\delta e$ and $\delta n_J$ have no meanings
when the stability conditions are employed 
because they require $\delta e = \delta n_J = 0$.
The constitutive equations are in the form of linear response
to the thermodynamic force
such as $\nabla_\mu u^\mu$ or $\nabla ^{\langle \mu} u^{\nu \rangle}$.
Note here that first order theory is independent
of the specific form of the distribution $f^i$.

These first order expressions
are acausal and unstable \cite{HiscockLindblom}
because they allow instantaneous propagation, and
one has to consider the second order corrections
to $s^\mu$ to obtain a causal theory.
The expansion of the entropy current at the second order yields
\begin{eqnarray}
\label{eq:ds2}
\delta s^\mu_{(2)} = -\frac{1}{2}\sum _i \int \frac{g_i d^3 p}{(2\pi )^3 E_i} p_i^\mu \frac{\delta f^{i2}}{f_0^i (1\pm f_0^i)} ,
\end{eqnarray}
which is negative when contracted with the flow $u_\mu$, 
representing the fact that system is in maximum entropy state.
This is the thermodynamic stability condition at the second order.
Its derivative $\partial_\mu \delta s^\mu_{(2)}$
obviously requires the information of $\delta f^i$.
This suggests that, unlike first order theory,
second order theory in general
depends on how the distribution is estimated.

\section{Second Order Distortion of Distribution}
\label{sodod}

The expansion of the phase space distribution $f^i = [\exp{(y^i)} \mp 1]^{-1}$ around the local equilibrium distribution $f_0^i$ up to the second order yields the distortion of the distribution 
\begin{equation}
\delta f^i = - f_0^i (1 \pm f_0^i) \delta y^i + \frac{1}{2} f_0^i (1 \pm f_0^i) (1 \pm 2 f_0^i) \delta y^{i2}.
\end{equation}
We estimate $\delta y^i$ in the extended moment method as $\delta y^i = p_i^\mu \sum_J q_i^J \varepsilon^J_\mu + p_i^\mu p_i^\nu \varepsilon_{\mu \nu}$, and the distortion up to the second order is expressed as
\begin{eqnarray}
\delta f^i &=& - f_0 (1 \pm f_0) (p_i^\mu \sum_J q_i^J \varepsilon^J_\mu + p_i^\mu p_i^\nu \varepsilon_{\mu \nu}) \nonumber \\
&+& \frac{1}{2} f_0 (1 \pm f_0) (1 \pm 2 f_0) (p_i^\mu p_i^\nu \sum_{J,K} q_i^J q_i^K \varepsilon^J_\mu  \varepsilon^K_\nu \nonumber \\
&+& 2 p_i^\mu p_i^\nu p_i^\rho \sum_J q_i^J \varepsilon^J_\mu \varepsilon_{\nu \rho} + p_i^\mu p_i^\nu p_i^\rho p_i^\sigma \varepsilon_{\mu \nu} \varepsilon_{\rho \sigma}).
\end{eqnarray}
In principle, the $10+4N$ unknowns are again determined by matching the macroscopic dissipative currents with the ones calculated in kinetic theory. Unlike in first order theory, though, the distortion involves higher order contributions in terms of the dissipative currents. The resultant $\varepsilon^J_\mu$ and $\varepsilon_{\mu \nu}$ would be, by omitting the third and higher order contributions, expressed as follows:
\begin{eqnarray}
\varepsilon_{\mu \nu} &=& \bigg( B_\Pi \Pi + B_{\delta e} \delta e + \sum _J B_{\delta n_J} \delta n_J \bigg) \Delta _{\mu \nu} \nonumber \\
&+& \bigg(\tilde{B}_\Pi \Pi + \tilde{B}_{\delta e} \delta e + \sum _J \tilde{B}_{\delta n_J} \delta n_J \bigg) u_\mu u_\nu \nonumber \\
&+& 2 B_W u_{( \mu} W_{\nu )} + 2 \sum_J B_{V_J} u_{( \mu} V^J_{\nu )} + B_\pi \pi_{\mu \nu} \nonumber \\
&+& \bigg(\beta_{\Pi \Pi} \Pi ^2 + \beta_{\delta e \delta e} \delta e ^2 + \sum_{J,K} \beta_{\delta n_J \delta n_K} \delta n_J \delta n_K \nonumber \\
&+& \beta_{\Pi \delta e} \Pi \delta e + \sum_J \beta_{\Pi \delta n_J} \Pi \delta n_J + \sum_J \beta_{\delta e \delta n_J} \delta e \delta n_J \nonumber \\
&+& \beta_{W W} W^\rho W_\rho + \sum_J \beta_{W V_J} W^\rho V^J_\rho \nonumber \\
&+& \sum_{J,K} \beta_{V_J V_K} V_J^\rho V^K_\rho + \beta_{\pi \pi} \pi^{\rho \sigma} \pi_{\rho \sigma} \bigg) \Delta_{\mu \nu} \nonumber \\
&+& \bigg( \tilde{\beta}_{\Pi \Pi} \Pi ^2 + \tilde{\beta}_{\delta e \delta e} \delta e ^2 + \sum_{J,K} \tilde{\beta}_{\delta n_J \delta n_K} \delta n_J \delta n_K \nonumber \\
&+& \tilde{\beta}_{\Pi \delta e} \Pi \delta e + \sum_J \tilde{\beta}_{\Pi \delta n_J} \Pi \delta n_J + \sum_J \tilde{\beta}_{\delta e \delta n_J} \delta e \delta n_J \nonumber \\
&+& \tilde{\beta}_{W W} W^\rho W_\rho + \sum_J \tilde{\beta}_{W V_J} W^\rho V^J_\rho \nonumber \\
&+& \sum_{J,K} \tilde{\beta}_{V_J V_K} V_J^\rho V^K_\rho + \tilde{\beta}_{\pi \pi} \pi^{\rho \sigma} \pi_{\rho \sigma} \bigg) u_\mu u_\nu \nonumber \\
&+& 2 \beta_{\Pi W} \Pi W_{(\mu} u_{\nu )} + 2 \sum_J \beta_{\Pi V_J} \Pi V^J_{(\mu} u_{\nu )} \nonumber \\
&+& 2 \beta_{\delta e W} \delta e W_{(\mu} u_{\nu )} + 2 \sum_J \beta_{\delta e V_J} \delta e V^J_{(\mu} u_{\nu )} \nonumber \\
&+& 2 \sum_J \beta_{\delta n_J W} \Pi W_{(\mu} u_{\nu )} + 2 \sum_{J,K} \beta_{\delta n_J V_K} \delta n_J V^K_{(\mu} u_{\nu )} \nonumber \\
&+& 2 \beta_{W \pi} W^\rho \pi_{\rho (\mu} u_{\nu )} + 2 \sum_J \beta_{V_J \pi} V_J^\rho \pi_{\rho (\mu} u_{\nu )} \nonumber \\
&+& \beta_{\Pi \pi} \Pi \pi_{\mu \nu} + \beta_{\delta e \pi} \delta e \pi_{\mu \nu} + \sum_J \beta_{\delta n_J \pi} \delta n_J \pi_{\mu \nu} \nonumber \\
&+& \bar{\beta}_{W W} W_\mu W_\nu + \sum_J \bar{\beta}_{W V_J} W_\mu V^J_\nu \nonumber \\
&+& \sum_{J,K} \bar{\beta}_{V_J V_K} V^J_\mu V^K_\nu + \bar{\beta}_{\pi \pi} \pi_{\mu}^{\ \rho} \pi_{\nu \rho} ,
\label{eq:et2}
\end{eqnarray}
\begin{eqnarray}
\varepsilon^J_\mu &=& \bigg( D^J_\Pi \Pi + D^J_{\delta e} \delta e + \sum _{K} D^J_{\delta n_{K}} \delta n_{K} \bigg) u_\mu \nonumber \\
&+& D^J_W W_\mu + \sum_{K} D^J_{V_{K}} V^{K}_\mu \nonumber \\
&+& \bigg( \delta^J_{\Pi \Pi} \Pi ^2 + \delta^J_{\delta e \delta e} \delta e^2 + \sum_{K,L} \delta^J_{\delta n_K \delta n_L} \delta n_K \delta n_L \nonumber \\
&+& \delta^J_{\Pi \delta e} \Pi \delta e + \sum_{K} \delta^J_{\Pi \delta n_K} \Pi \delta n_K + \sum_{K} \delta^J_{\delta e \delta n_K} \delta e \delta n_K \nonumber \\
&+& \delta^J_{W W} W^\nu W_\nu + \sum _{K} \delta^{J}_{W {V_K}} W^\nu V^K_\nu \nonumber \\
&+& \sum _{K,L} \delta^J_{V_K V_L} V_K^\nu V^L_\nu \bigg) u_\mu + \delta^J_{\Pi W} \Pi W_\mu + \delta^J_{\delta e W} \delta e W_\mu \nonumber \\
&+& \sum_K \delta^J_{\delta n_K W} \delta e W_\mu + \sum_K \delta^J_{\Pi V_K} \Pi V^K_\mu \nonumber \\
&+& \sum_K \delta^J_{\delta e V_K} \delta e V^K_\mu + \sum_{K,L} \delta^J_{\delta n_K V_L} \delta n_K V^L_\mu \nonumber \\
&+& \delta^J_{W \pi} W^\nu \pi_{\mu \nu} + \sum_K \delta^J_{V_K \pi} V_K^\nu \pi_{\mu \nu} ,
\label{eq:ev2}
\end{eqnarray}
where $\delta$'s and $\beta$'s are the second order prefactors which can be calculated in kinetic theory. 
These new terms might add higher order contributions on the left hand side of the second order constitutive equations (\ref{eq:Pi})-(\ref{eq:pi}) which can be absorbed into the right hand side by using the Navier-Stokes expressions for the dissipative currents $\Pi$, $W^\mu$, $V_J^\mu$ and $\pi^{\mu \nu}$.

\section{Multi-Component Systems with Single Conserved Current}
\label{mcc}

If the system has only one conserved current, then the number of dissipative currents match that of unknowns within the framework of the conventional Grad's 14-moment method even if inelastic collisions are present. Then we have
\begin{eqnarray}
\label{eq:deltay_alt}
\delta y^i &=& p_i^\mu \varepsilon _{\mu} + p_i^\mu p_i^\nu \varepsilon _{\mu \nu} ,
\end{eqnarray}
instead of Eq.~(\ref{eq:dym}). The expression for the entropy production is modified to
\begin{eqnarray}
\label{eq:ent_pro_IS2_B}
\partial _\mu s^\mu &=& \varepsilon_{\mu \nu} \partial _\alpha I^{\mu \nu \alpha} \geq 0 ,
\end{eqnarray}
which means the second moment charge-weighted moment equations $\partial _\alpha I_J^{\mu \alpha} = Y_J^\mu$ cannot be constrained from the second law of thermodynamics. If we assume that $Y_J^\mu$ is a macroscopic quantity which can be expressed as a linear combination of all the dissipative currents, we obtain the same constitutive equations as Eqs.~(\ref{eq:Pi})-(\ref{eq:pi}) with different transport coefficients. For the scalar, the vector and the tensor dissipative currents, Onsager's reciprocal relations can be satisfied if we are allowed to adjust the transport coefficient matrices $\mathcal{C}_i$'s and
if the semi-positive definiteness of the resulting transport coefficient matrices $\mathcal{A}_i$ is preserved. 
Since this formalism is not equivalent to the multiple conserved current theory we developed, it should be investigated which formalism we should follow to discuss the single conserved current systems.


\end{document}